\documentclass[fleqn,usenatbib,useAMS]{mnras}

\usepackage{natbib}

\usepackage{ragged2e}
\usepackage{graphicx}
\usepackage{color}
\usepackage{txfonts}
\usepackage{soul}
\usepackage{lscape,supertabular,longtable,pdflscape}
\usepackage{wrapfig}
\usepackage{subcaption}
\usepackage{cleveref}
\usepackage{bbding}

\newcommand{\msunyr}{\ensuremath{\mathrm{M}_{\odot}{\rm yr}^{-1}}}   
\newcommand{\msun}{\ensuremath{\mathrm{M}_{\odot}}}   
\newcommand{\mini}{\ensuremath{M_{\rm ini}}}                         
\newcommand{\lsun}{\ensuremath{\mathit{L}_{\odot}}}                  

\newcommand{\vcrit}{\ensuremath{\upsilon_{\rm crit}}}                         
\newcommand{\vini}{\ensuremath{\upsilon_{\rm ini}}}                         

\newcommand{\mdot}{\ensuremath{\dot{M}}}                             
\newcommand{\mstar}{\ensuremath{\mathit{M}_{\star}}}                 
\newcommand{\teff}{\ensuremath{\mathit{T}_{\rm eff}}}                

\newcommand{\logteff}{\ensuremath{\mathit{\log(\teff)}}}              
\newcommand{\logl}{\ensuremath{\mathit{\log(L/\lsun)}}}               

\newcommand{\nitrgn}{\ensuremath{^{14}\mathrm{N}} }              

\definecolor{green}{rgb}{0.3,0.7,0.}

\definecolor{green}{rgb}{0.3,0.7,0.}

\definecolor{orange}{rgb}{0.5,0.2,0.0}

\hypersetup{draft}
\begin{document}



\title[Stellar Models with Rotation: $Z=0$]{Grids of stellar models with rotation: V. Models from 1.7 to 120$\,\msun$ at zero metallicity}

\author[L.J.Murphy et al.]{Laura~J.~Murphy$^{1}$\thanks{E-mail: murphl25@tcd.ie}, 
Jose H. Groh$^{1}$,
Sylvia Ekstr\"om$^{2}$,
Georges Meynet$^{2}$,
\newauthor
Camila Pezzotti$^{2}$,
Cyril Georgy$^{2}$,
Arthur Choplin$^{3}$,
Patrick Eggenberger$^{2}$,
Eoin Farrell$^{1}$,
\newauthor
Lionel Haemmerl\'e$^{2}$,
Raphael Hirschi$^{4}$,
Andr\'e Maeder$^{2}$,
Sebasti\'en Martinet$^{2}$
 \\
 $^{1}$School of Physics, Trinity College Dublin, the University of Dublin, College Green, Dublin\\
 $^{2}$Department of Astronomy, University of Geneva, Chemin des Maillettes 51, 1290 Versoix, Switzerland\\
 $^{3}$Institut d’Astronomie et d’Astrophysique, Université Libre de Bruxelles (ULB), CP 226, 1050 Brussels, Belgium\\
 $^{4}$Astrophysics Group, Keele University, Keele, Staffordshire ST5 5BG, UK
 }

\date{Accepted XXX. Received YYY; in original form ZZZ}

\pubyear{2020}


\label{firstpage}
\pagerange{\pageref{firstpage}--\pageref{lastpage}}
\maketitle

\begin{abstract}
   {Understanding the nature of the first stars is key to understanding the early universe. With new facilities such as JWST we may soon have the first observations of the earliest stellar populations, but to understand these observations we require detailed theoretical models.}
   {Here we compute a grid of stellar evolution models using the Geneva code with the aim to improve our understanding of the evolution of zero-metallicity stars, with particular interest in how rotation affects surface properties, interior structure, and metal enrichment.}
   {We produce a range of models of initial masses ($\mini$) from $1.7\,\msun$ to $120\,\msun$, focusing on massive models of $9\,\msun \leq \mini \leq 120\,\msun$. Our grid includes models with and without rotation, with rotating models having an initial velocity of 40\% of the critical velocity.}
  {We find that rotation strongly impacts the evolution of the first stars, mainly through increased core size and stronger H-burning shells during core He-burning. Without radiative mass loss, angular momentum builds at the surface in rotating models, thus models of initial masses $\mini \geq 60 \, \msun$ reach critical rotation on the main sequence and experience mass loss. We find that rotational mixing strongly affects metal enrichment, but does not always increase metal production as we see at higher metallicities. This is because rotation leads to an earlier CNO boost to the H shell during He-burning, which may hinder metal enrichment depending on initial mass and rotational velocity. Electronic tables of this new grid of Population III models are publicly available.}
  
\end{abstract}


\section{Introduction}\label{sec:intro}

The first stars formed from metal-free primordial gas at a redshift of $z\approx20-30$, a few hundred million years after the Big Bang \citep{Bromm2013FormationStars}. This first generation of stars contributed significant amounts of ionising photons to the reionisation of the universe, and provided the first heavy elements, fueling later generations of higher metallicity stars. The interaction of the first stars with the interstellar medium impacted the universe extensively, in particular through ionisation \citep{BarkanaLoeb2001,Whalen2004,Kitayama2004,Alvarez2006,Wyithe2007TheHydrogen,Whalen2008,Wise2008}, chemical enrichment \citep{Mackey2003,Kitayama2005,Greif2007,Whalen2008enrichment,Joggerst2010,HegerWoosley2010,Greif2010,Kobayashi2011,Hartwig2018,Hartwig2019,Chiaki2019,Welsh2019,Hicks2020,Magg2020}, as well as their explosive deaths as supernovae (SNe) \citep{Umeda2002,Nozawa2003,Cayrel2004,Tominaga2007}. Studying these primordial stars can tell us about the nature of the earliest explosions in the universe and allow us to investigate their progenitors. The nature of these first stars not only impacts the properties of their SNe, but also the amount of ionising flux at high redshift \citep{Schaerer2002,Wyithe2003WasStars}, the formation of supermassive black holes \citep{Sakurai2013FormationAccretion,Haemmerle2018,Smith2018,ReganDownes2018,Woods2019,Haemmerle2020}, and the rates of gravitational wave signals from primordial neutron star and black hole mergers \citep{Kinugawa2014,Kinugawa2016a,Kinugawa2016b,Kinugawa2017}. 

Current stellar evolution models indicate that the evolution of the first stars and fate of the first explosions at high redshift is mainly determined by initial mass and processes such as convection, rotation, and magnetic fields \citep{Marigo2001Zero-metallicityMass,Ekstrom2008EffectsStars,Yoon2012EvolutionFields}. There is large uncertainty on all these properties, and even more so at zero metallicity where we have no direct observational constraints. The first hydrodynamical simulations (e.g. \citealt{Abel2002}) predicted preferential formation of very massive first stars ($\geq \! 100 \,\msun$; \citealt{Bromm2002TheCloud}), which would explode as pair instability SNe (PISNe). However, more recent simulations predict significant fragmentation \citep{Stacy2010,Clark2011TheProtostars}, the formation of binaries \citep{Turk2009}, and a wide initial mass distribution from tens to hundreds of solar masses \citep{Hirano2014,Hirano2015}. This implies that many of them would explode as core-collapse SNe (CCSNe) and not as PISNe, which is a significant difference, mainly because CCSNe leave behind a black hole or a neutron star, while PISNe leave no remnant. No clear signatures of chemical enrichment from PISNe of extremely metal-poor stars have been seen  \citep{Umeda2002,Nomoto2006} with \citet{Karlsson2008} predicting a number fraction of primordial PISNe of less than 0.07, suggesting that Population III (Pop~III) stars with masses above $100\,\msun$ may have been rare\footnote{There are a few candidates for PISNe at high redshift \citep{GalYam2009,Cooke2012}, however they are not thought to arise from Pop~III stars. Furthermore, their identification as PISNe is disputed \citep{Dessart2013}.}. This is in agreement with \citet{Bromm2013FormationStars} which suggests that fragmentation and the formation of multiple star systems move away from the idea of a predominantly very massive initial mass function (IMF) as initially predicted in \citet{Bromm2002TheCloud}. Although there is general agreement on a top-heavy primordial IMF \citep{Stacy2016}, there is still little constraint on its exact distribution, so theoretical models must continue to consider a large range of initial masses.

Several groups have investigated the evolution of the first stars with numerical stellar evolution codes. \citet{Marigo2001Zero-metallicityMass} studied zero-metallicity, non-rotating stellar evolution models with initial masses in the range $\mini\!=\!0.7-100\,\msun$, with subsequent work focusing on rotating models with $\mini\!=\!120-1000\,\msun$ \citep{Marigo2003}. They were particularly interested in how the lack of CNO group elements in these models changes their nuclear energy generation and interior structure. \citet{Marigo2003} also explored the effect of mass loss for models of higher initial mass, finding that the radiation pressure in these stars is not an efficient driving force of mass loss. These authors also discussed that rotation could trigger mass loss, but the star would then spin down quickly afterwards. While \citet{Marigo2003} presented a much improved understanding of the effect of rigid rotation on primordial stellar evolution, there was still a need to investigate the effects of differential rotation. This was addressed by \citet{Ekstrom2008EffectsStars} where primordial stars of initial masses $\mini\!=\!9-200\,\msun$ were investigated using the Geneva stellar evolution code (GENEC), and advection was included in the angular momentum transport, allowing for a more accurate treatment of rotational mixing. These authors found that, although still impactful, the effect of rotation on the first stars is smaller than for even extremely metal-poor stars. This was believed to result from weak meridional circulation, and continuous nuclear burning at the end of the main sequence (MS).

\citet{Yoon2012EvolutionFields} included magnetic fields in Pop~III stellar models assuming a Taylor-Spruit dynamo, for an initial mass range $\mini\!=\!10-1000\,\msun$. This allowed for a different approach to the effects of rotation in these stars where chemically homogeneous evolution (CHE) is achieved. \citet{Yoon2012EvolutionFields} also predicted the final fates of these stars in a parameter space spanned by initial mass and initial rotation (see their Fig. 12). It was found that CHE is favoured in a higher mass star, until at a certain mass when mass loss due to critical rotation hinders chemical mixing. This is important because CHE has a significant effect on the final fates of these stars, leading to more explosive phenomena such as gamma-ray bursts, hypernovae and Type Ibc PISNe. This work also showed that slower rotators, which do not achieve CHE, end their lives either as Type-II SNe or by collapsing to a black hole. More recently, \citet{Windhorst2018OnTransits} produced a grid of Pop~III models with MESA \citep{Paxton2011,Paxton2013,Paxton2015} and investigated their detectability through cluster caustic transits. 

Stellar evolution models of He cores were used to investigate the death of massive Pop~III stars in \citet{Heger2002}. Through studying a range of He cores of masses from 60-140$\,\msun$ the nucleosynthesis yields were determined for stars expected to undergo pulsational pair instabilities. This was important for determining the chemical signature of the first stars and their impact on the metallicity of subsequent stellar populations.

There have been many developments in the field in recent years creating a need for updated theoretical models of primordial stars. There has been more research on observations of second generation (Pop~II) stars and how differences in assumptions of the first stars would affect these observational signatures. \citet{sarmento2019} varied the IMF and the critical metallicity defining the boundary between Pop~III and Pop~II stars investigating the SN yields from the Pop~III models with observations of Pop~II stars. They concluded that the Pop~III IMF is dominated by stars in the mass range $\mini\!=\!20-120\,\msun$ that generate SN with carbon-enhanced spectra. Formation scenarios of extremely metal-poor (EMP) stars were presented in \citet{Hartwig2018fingerprintpopIII}, and used to investigate their enrichment from Population III stars. There has also been much research on the topic of carbon-enhanced metal-poor (CEMP) stars \citep{Limongi2003,Umeda2003,Meynet2006cemp,Choplin2016}. These are of particular interest in relation to the early universe as they shed light on possible constraints for the first generations of stars \citep{HegerWoosley2010,Tominaga2014,Takahashi2014,Choplin2017paper,Choplin2017letter,Choplin2018paper,Choplin2020proceedings}, particularly their mixing processes and rotation. 
Mixing of the Helium and Hydrogen layers in the first generation of stars could reproduce the abundance pattern of CEMP-no stars \citep{Choplin2016} i.e. no enhancement from s-process and r-process elements. Moreover, interactions between He rich and H rich layers in Pop~III stars give rise to different reaction chains and affect final abundances, with interesting implications for CEMP-no stars \citep{Clarkson2018,Clarkson2020}. These stars can therefore help us to understand how the first stars may have influenced the generations that followed them. 

Ahead of the launch of the James Webb Space Telescope (JWST) there is increased research on the observability of Pop~III stars \citep{zackrisson2015observepopIIIjwst,Windhorst2018OnTransits} and the likelihood to observe the first stellar populations. Despite previous and ongoing surveys a metal-free star is yet to be observed \citep{Beers1992,1995McWilliam,Ryan1996,Cayrel2004,Christlieb2008,Roederer2014,Howes2016,Starkenburg2017}. However, recent studies have addressed the detectability of SNe and PISNe from the early universe \citep{Whalen2013SeeingJWST,Whalen2013CCSNe,Whalen2013PISNe,Whalen2014PPISNe}, suggesting that these events will be easily observed by the next generation of optical space telescopes up to a redshift $z = 15 - 30$ \citep{Tanaka2013,deSouza2013,deSouza2014,Moriya2019}. 

In addition, substantial developments have occurred in the area of gravitational-wave research in recent years, with  observations of black hole mergers and neutron star mergers \citep{LIGO2017,Abbott2019}. This has created a surge in research related to Pop~III progenitors of black holes. Studies such as \citet{Latif2013}, \citet{Regan2017} and \citet{Johnson2019} modelled the formation of direct collapse black holes from primordial gas, \citet{Hartwig2018GWs} investigated the merger rates of black holes at high redshifts ($z\geq\!15$), and \citet{Uchida2019BHgravwaves} computed the gravitational waves that would be emitted by the collapse of a $320\,\msun$ Pop~III star. Furthermore, recent studies have provided new insights into the nature of Pop~III supermassive stars and their role in the formation of the first quasars \citep{Hosokawa2013,Umeda2016,Woods2017,Haemmerle2018}. This work has shown that Pop~III protostars can continue to accrete material at rates of $\geq 0.01\,\msunyr$ towards masses of $\geq\!\! 10^5\,\msun$ without an increase in ionising feedback that would halt accretion, which favours the case for the formation of the first quasars through direct collapse. The recent discovery of the black hole merger event GW190521 \citep{ligo20report,ligo20implications} has particularly drawn attention to Pop~III stellar evolution, given that these stars offer explanation for the mass of this merger, which falls within the so-called pair instability mass gap. Pop~III stars have been discussed as progenitors for this event in \citet{Farrell2020_GW190521}, \citet{Kinugawa2020}, \citet{Liu2020}, and \citet{Safarzadeh2020} with \citet{Tanikawa2020} also proposing that Pop~III binary black holes may fall within the mass gap. With renewed interest in the study of these primordial stars, it is imperative that we have updated and detailed theoretical models to support this new research.

Since these zero-metallicity stars were studied in \citet{Ekstrom2008EffectsStars}, a new series of papers was started to investigate stellar evolution with the Geneva code. These publicly-available grids include updated physical ingredients and are suitable for various metallicities. \citet{Ekstrom2012GridsZ=0.014} (hereafter Paper I, $Z=0.014$) investigated solar metallicity models. This was followed by grids at lower metallicities in \citet{Georgy2013Grids0.002} (hereafter Paper II, $Z=0.002$) and \citet{Groh2019Grids0.0004} (hereafter Paper III, $Z=0.0004$). In this paper we present a new grid of Geneva stellar evolution models at $Z=0$. Our paper is organised as follows. We present our new stellar models in \Cref{sec:ingredients} and discuss their evolution in the Hertzsprung-Russell (HR) diagram in \Cref{sec:HRDevol}. In \Cref{sec:roteffects} we discuss the effects of rotation on Pop~III stars, while \Cref{sec:criticalrot} investigates the possibility of mass loss in fast-rotating Pop~III stars. We present the metal yields from our models in \Cref{sec:enrichment} and discuss the effects of changing the initial rotation in \Cref{change_vel}. \Cref{conclusions} presents our concluding remarks.

\section{Stellar models: physical ingredients and electronic tables}\label{sec:ingredients}

We use the latest version of the Geneva stellar evolution code, GENEC, to compute zero-metallicity models with a primordial initial composition of $X\!=\!0.7516$, $Y\!=\!0.2484$, and $Z\!=\!0$. The physical ingredients of the models presented in this work are consistent with those described in Papers I, II and III of higher metallicities \citep{Ekstrom2012GridsZ=0.014,Georgy2013Grids0.002,Groh2019Grids0.0004} and we refer the reader to these papers for full details. This includes the opacities which are generated using the OPAL tool (based on \citealt{IglesiasRogers1996}) and are complemented at low temperatures by opacities from \citet{ferguson2005}, and the nuclear reaction rates which are taken mainly from the Nacre database \citep{angulo1999}. Convective zones are determined using the Schwarzschild criterion, and for the MS and the He-burning phase the convective core is extended with an overshoot parameter $d_{\rm over}/H_P=0.1$, where $d_{\rm over}$ is the distance of overshooting beyond the Schwarzchild boundary and $H_P$ is the pressure scale-height at the edge of the core. We use a value of $d_{\rm over}/H_P=0.1$ for consistency with Papers I, II and III. 
It has been found in recent research however, that the overshooting parameter could be higher for massive stars, with $d_{\rm over}/H_P=0.3-0.5$ matching some observations of massive MS stars \citep{Castro2014overshoot,Schootemeijer2019,HigginsVink2019}.

Our grid consists of models at zero metallicity in the mass range $1.7 \, \msun \leq \mini \leq 120 \, \msun$. We compute non-rotating and rotating models, the latter with an initial equatorial rotational velocity of $\vini=0.4 \, \vcrit$, where  $\vcrit \!\!=\!\! \sqrt{\frac{2}{3}\frac{GM}{R_{\rm pol,crit}}}$ is the break-up velocity at critical rotation and $R_{\rm pol,crit}$ is the polar radius at $\vcrit$. The models with $\vini=0.4 \, \vcrit$ are consistent with Papers I, II and III, and based on the peak velocity distribution of young solar-metallicity B-type stars in \citet{Huang2010}. We have also produced models at the slower rotational velocity of $\vini=0.2 \, \vcrit$ for certain initial masses. This allows us to investigate the impact of a change of the initial rotation on the results, rather than relying on rotating models at a single velocity versus non-rotators. 

Recent work indicates that extremely low metallicity stars may have rotated as fast as $\vini=0.7\vcrit$ in order to reproduce the abundance pattern of some CEMP stars enriched in s-process elements \citep{Choplin2017letter,Choplin2020proceedings}. Simulations of Pop~III star formation \citep{Stacy2011rotation,Stacy2013rotation} also indicate that these early stars would have formed with significant rotation. Since these fast rotating models are challenging to compute due to convergence problems, we defer to future work an extension of our grid to models with higher initial surface rotation. 

We note that we do not consider the effects of magnetic fields in this work, and so our models are differentially rotating. The treatment of rotation follows that of Papers I, II and III, having been developed in a series of papers by the Geneva group \citep{maeder1997shear,meynetmaeder1997,maederzahn1998medvel,maedermeynet2000mdot}. The diffusion coefficients follow the \citet{zahn1992} and \citet{maeder1997shear} prescriptions for horizontal and shear diffusion respectively, and the treatment of advection follows the \citet{zahn1992} prescription.

Our models predict that Pop~III stars in the range 9-120$\,\msun$ are hot stars throughout their lifetimes. 
Theoretical works suggest decreasing mass loss with metallicity \citep[e.g.][]{vink2001}, with zero or negligible mass loss at $Z=0$ \citep{Krticka2006,Krticka2009}. Therefore, our models have no mass loss except when approaching critical rotation. Upon reaching critical velocity the outer layers of the star become gravitationally unbound \citep{2011Krticka} so it is expected that some mass would be removed until the star spins down below the critical velocity.

It is numerically difficult to compute the models when the star is rotating at critical velocity. In general, GENEC calculates the amount of mass that should be removed in order to bring the model back below the critical limit (see also \citealt{Georgy2013mechmdot}). This is based on the amount of angular momentum that must be lost for the star to become subcritical again, given by 
\vspace{-0.15cm}
\begin{equation}
    \Delta \mathcal{L}_{\rm mec}=\Delta M_{\rm mec} R_E^2 \Omega_1~,
\end{equation}
where $\Delta M_{\rm mec}$ is the amount of mass lost during each time step in the disc, $R_E$ is the equatorial radius of the star, and $\Omega_1$ is the angular velocity of the first layer. Using this value, $\Delta M_{\rm mec}$, a mechanical mass loss rate can be imposed when the star reaches critical velocity. When using this implementation, the estimated value for $\Delta M_{\rm mec}$ brings the model just below the critical limit. However, unlike models at high metallicity, further mass loss through radiative winds is inefficient in Pop~III stars, and our zero-metallicity models remain close to the critical limit. This causes numerical problems. To successfully evolve the models, instead of using mechanical mass loss we assume an averaged mass loss rate of $\mdot=10^{-5} \,\msunyr$ upon reaching the critical limit. For computational convenience, the mass loss rate is kept at that value until the star is sufficiently far from the critical limit. Physically, this could correspond to another process such as pulsational mass loss. We note that this leads to a difference in the angular momentum profile depending on which  mass loss regime is imposed. GENEC's implementation assumes that material and angular momentum are lost at the equator which would lead to an equatorial decretion disk, while our mass loss regime assumes for simplicity a spherical distribution of angular momentum loss which would form circumstellar material (CSM). We encourage further hydrodynamical studies to explore the behavior of zero-metallicity stars near critical rotation.

Similarly to Papers I, II and III, electronic tables of the models are publicly available\footnote{See \href{https://obswww.unige.ch/Research/evol/tables\_grids2011/}{https://obswww.unige.ch/Research/evol/tables\_grids2011/}}. For each model, the evolutionary track consists of 400 selected data points, with each one corresponding to a given evolutionary stage. Points of different evolutionary tracks with the same number correspond to similar stages to facilitate interpolation of the  evolutionary tracks. The reader may refer to Paper I \citep{Ekstrom2012GridsZ=0.014} for details on the numbering of these points and their corresponding evolutionary phases. With this grid publicly available it can be used as input for computing interpolated tracks, isochrones, and population synthesis models using the Geneva tools\footnote{\href{https://www.unige.ch/sciences/astro/evolution/en/database/}{https://obswww.unige.ch/Recherche/evoldb/index/}}. A detailed description of the online tools is presented in \citet{Georgy2014SYCLIST}.


\section{Overall evolution of Pop~III stars on the HR Diagram}\label{sec:HRDevol}
\begin{figure*}
        \includegraphics[width=0.48\linewidth]{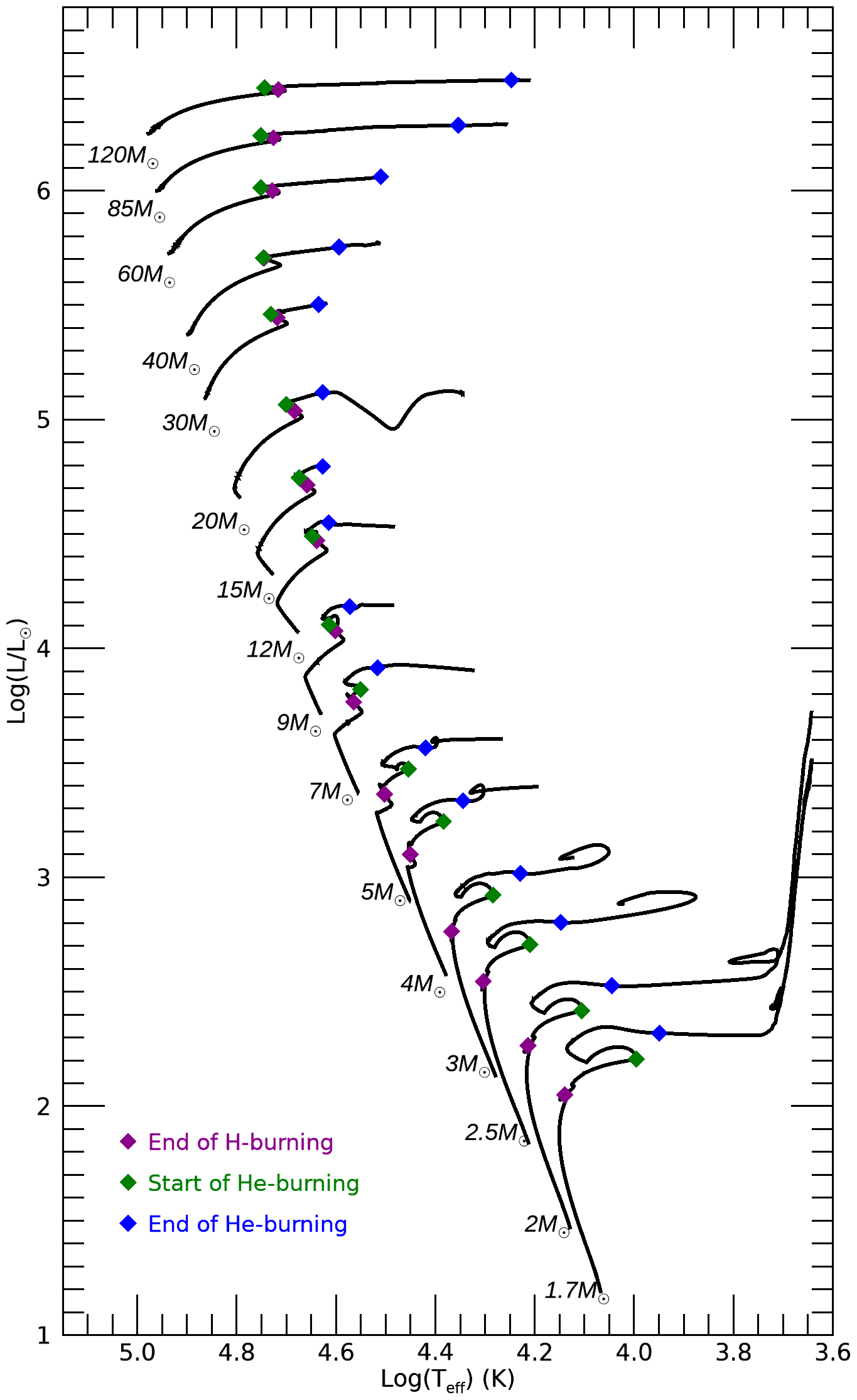}
        \includegraphics[width=0.48\linewidth]{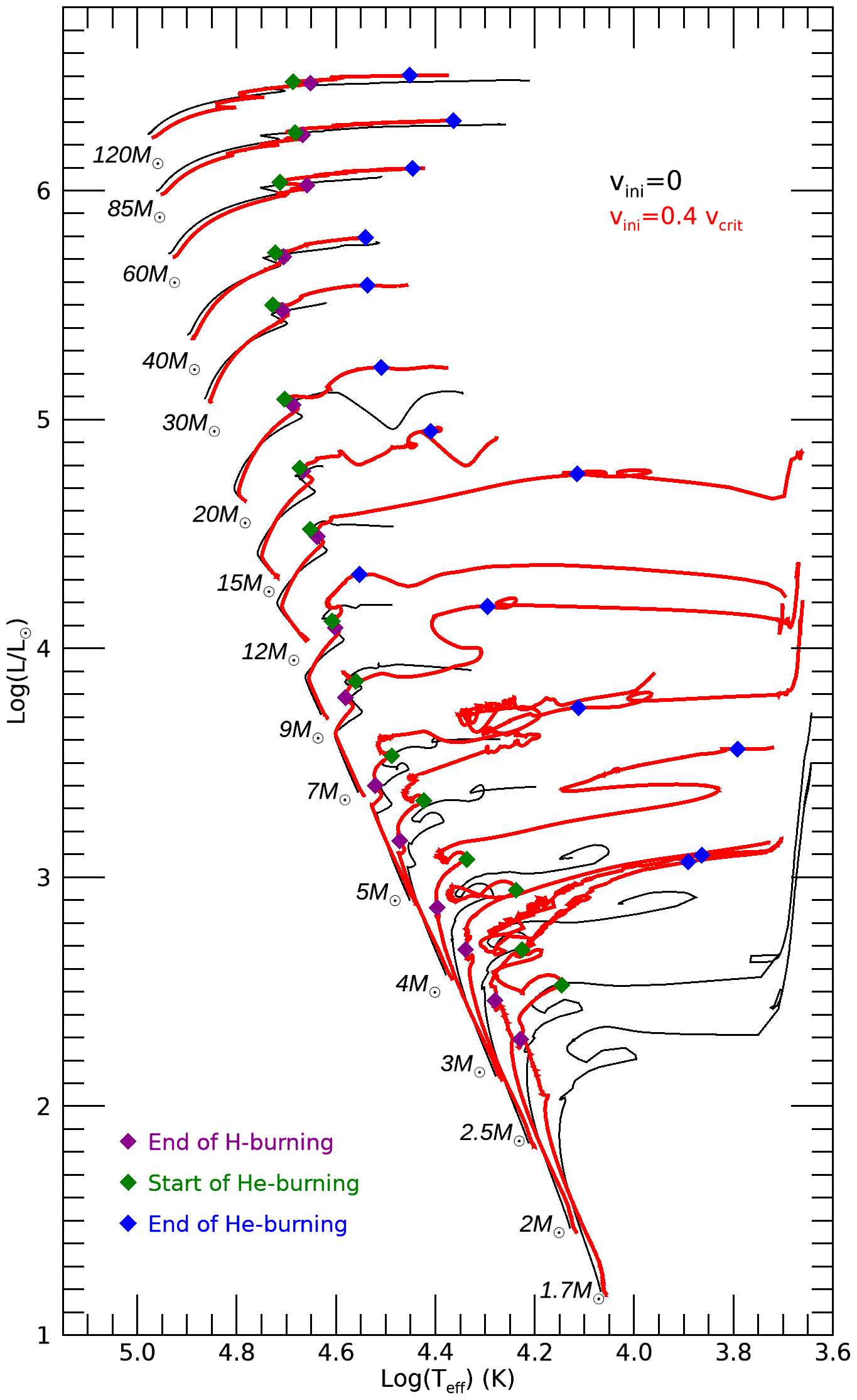}
    \caption{Left: Evolutionary tracks of models in the mass range $1.7\,\msun \leq \mini \leq 120\,\msun$ for non-rotating models (black). Right: Comparison between non-rotating (black) and rotating models with $\vini=0.4\,\vcrit$ (red). Key evolutionary stages are given in the legend. }
    \label{fig:HRD}
\end{figure*}

\begin{figure}
   \centering
   \includegraphics[width=\linewidth]{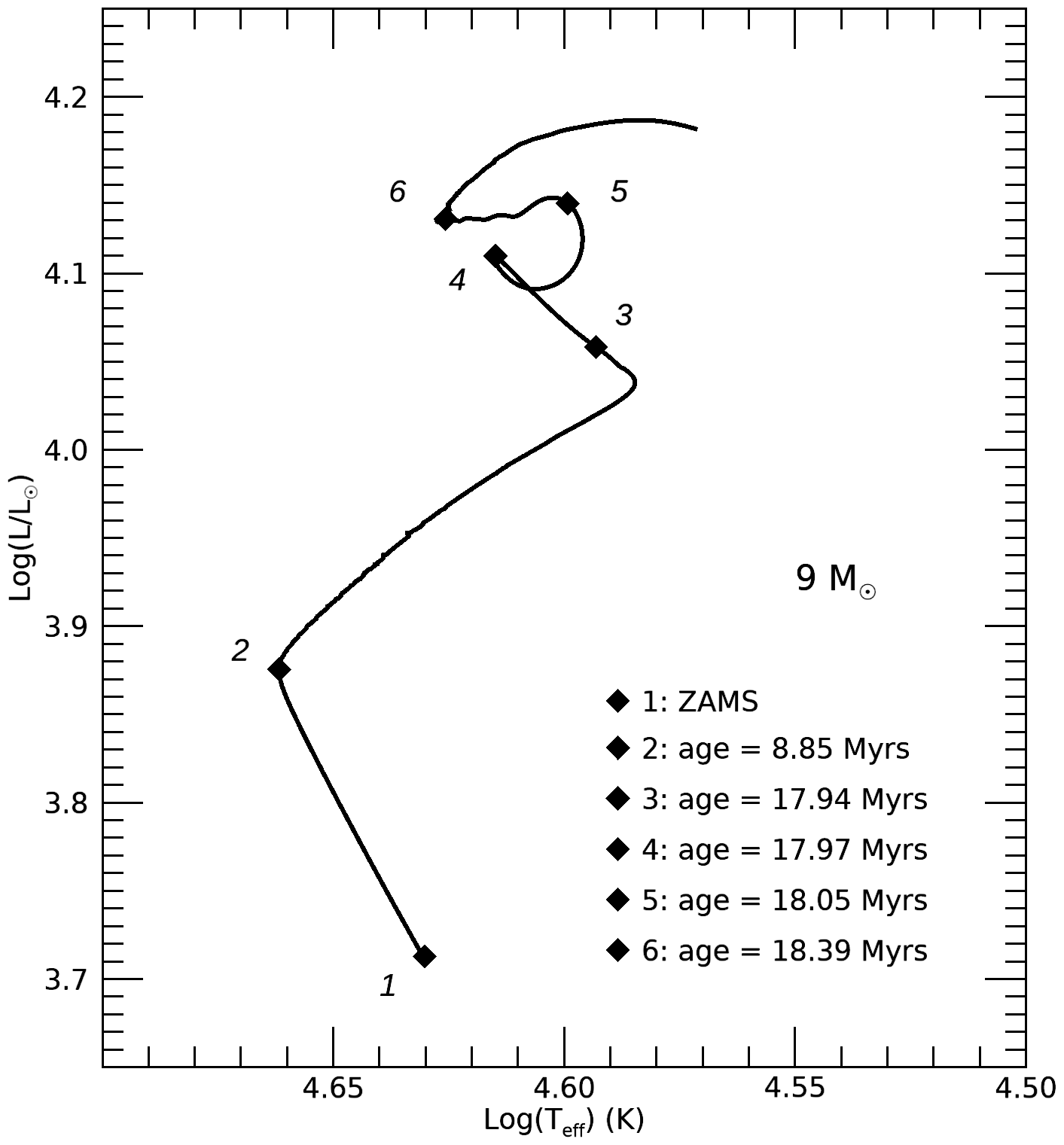}
      \caption{Evolution for the non-rotating $9\,\msun$ model. Selected key stages of the evolution are indicated. Stages 1 and 2 illustrate the contracting phase during H-burning where stage 1 marks the ZAMS and stage 2 marks where the CNO cycle becomes dominant, ie. $\epsilon_{\rm CNO} > \epsilon_{\rm pp}$. Stages 3-6 are used to understand the evolution from the late MS through to early He-burning, they correspond to the interior structure profiles in \Cref{fig:9centsurf,fig:9engen,fig:9egrav}.
              }
         \label{fig:9HR}
   \end{figure}

The stellar evolution tracks for both non-rotating and rotating models are shown in \Cref{fig:HRD}. As expected, the effective temperature (\teff) and luminosities at the zero-age main sequence (ZAMS) increase with increasing mass for both rotating and non-rotating models.  Considering the non-rotating models in the mass range $9\, \msun \! \leq  \mini \leq \! 120\, \msun$ first, models with $\mini \geq 30 \, \msun$ show a qualitative evolution during the MS that resembles that of higher metallicity models \citep{Ekstrom2012GridsZ=0.014,Georgy2013Grids0.002,Groh2019Grids0.0004}. In these models, the surface properties of the star during H-burning steadily evolve to higher luminosities and cooler surface temperatures as the stellar envelope expands. At the end of H-burning, stars with $\mini \geq 30 \, \msun$ will have cooled to approximately $\logteff = 4.7$. As seen in previous works \citep{Ekstrom2008EffectsStars}, models with $\mini=9-20\,\msun$ spend a significant fraction of their MS lifetime burning H with only proton-proton (p-p) chain reactions, since their central temperature is not yet high enough for producing C and O through the triple-alpha reaction.  The star keeps contracting until the CNO cycle begins, and this phase corresponds to the evolution towards higher \teff\ from the ZAMS (\Cref{fig:HRD} and stages 1-2 in \Cref{fig:9HR}). The length of time a model spends in this contracting p-p chain phase decreases with increasing initial mass. This is because models of higher initial mass have higher central temperatures and therefore produce CNO elements earlier in their evolution than less massive models. Models with $\mini \geq 30 \, \msun$ have a core that is hot enough to produce the C, N, O catalysts immediately and can burn H similarly to higher metallicity models.

Pop~III stars have smoother transitions between burning phases than high-metallicity stars \citep{Ekstrom2008EffectsStars,Marigo2001Zero-metallicityMass}. This is evident from the near overlap of end H-burning and start He-burning phases (\Cref{fig:HRD}). Non-rotating models in the range 9-20$\,\msun$ show a distinctive feature at the start of He-burning (loop next to the green point in \Cref{fig:HRD}, more clearly seen in \Cref{fig:9HR}) that are relevant since they leave imprints in the abundance profile and core size, which affects the subsequent evolution during He-burning, and in particular the final \teff. 

We use the non-rotating $9\,\msun$ model to illustrate the change in surface properties that gives rise to this distinctive feature (stages 4-6 in \Cref{fig:9HR}). This model shows a sharp decrease in $\teff$ immediately after He ignition (stages 4-5), followed by a gradual increase in $\teff$ (stages 5-6). When H is depleted in the core the continuing contraction of the star (stages 3-4) ignites the H shell leading to a boost in luminosity at the surface. He core burning then begins and there is a further boost to the luminosity (stages 4-5). Our models show that just prior to He core ignition when the H-burning shell dominates, the now inactive core is strongly contracting while the envelope expands due to the energy boost from this H shell (stages 4-5). This puts the star out of thermal equilibrium. When He core burning begins the star regains thermal equilibrium (stages 5-6). The combined effects of the H-burning shell, He core contraction, He core ignition, and the timescale to regain thermal equilibrium cause the complex \teff\ evolution at the transition from H to He-burning. We discuss this in further detail in \Cref{sec:app1}.

Once the star is in thermal equilibrium it evolves towards lower \teff\ during He core burning (\Cref{fig:HRD}). A clear trend with initial mass can be seen for the end He-burning position of non-rotating models on the HR diagram. For less massive models there is little change in $\teff$ during the He-burning phase, however, as initial mass increases it can be seen that they evolve to lower $\teff$ during this evolutionary phase. 

Also included in \Cref{fig:HRD} are the intermediate mass models in the mass range $1.7\,\msun \! \leq \! \mini \! \leq \! 7\,\msun$.  The non-rotating intermediate mass models (left panel of \Cref{fig:HRD}) spend even longer than the massive models in the contracting phase where only the p-p chain reactions contribute to the nuclear energy production. The distinctive loop feature at the beginning of the He-burning phase is also prominent in these models. Notably, none of the intermediate mass non-rotators become red giants before the end of the core He-burning phase. While the key focus of this paper is the effect of rotation on massive Pop~III models, these lower mass models complement our grid and will be very useful for future work, for example in population synthesis.


\section{Effects of Rotation on Pop~III stars}\label{sec:roteffects}

\subsection{HR diagram} \label{subsec:HR}
\begin{figure}
   \centering
   \includegraphics[width=\linewidth]{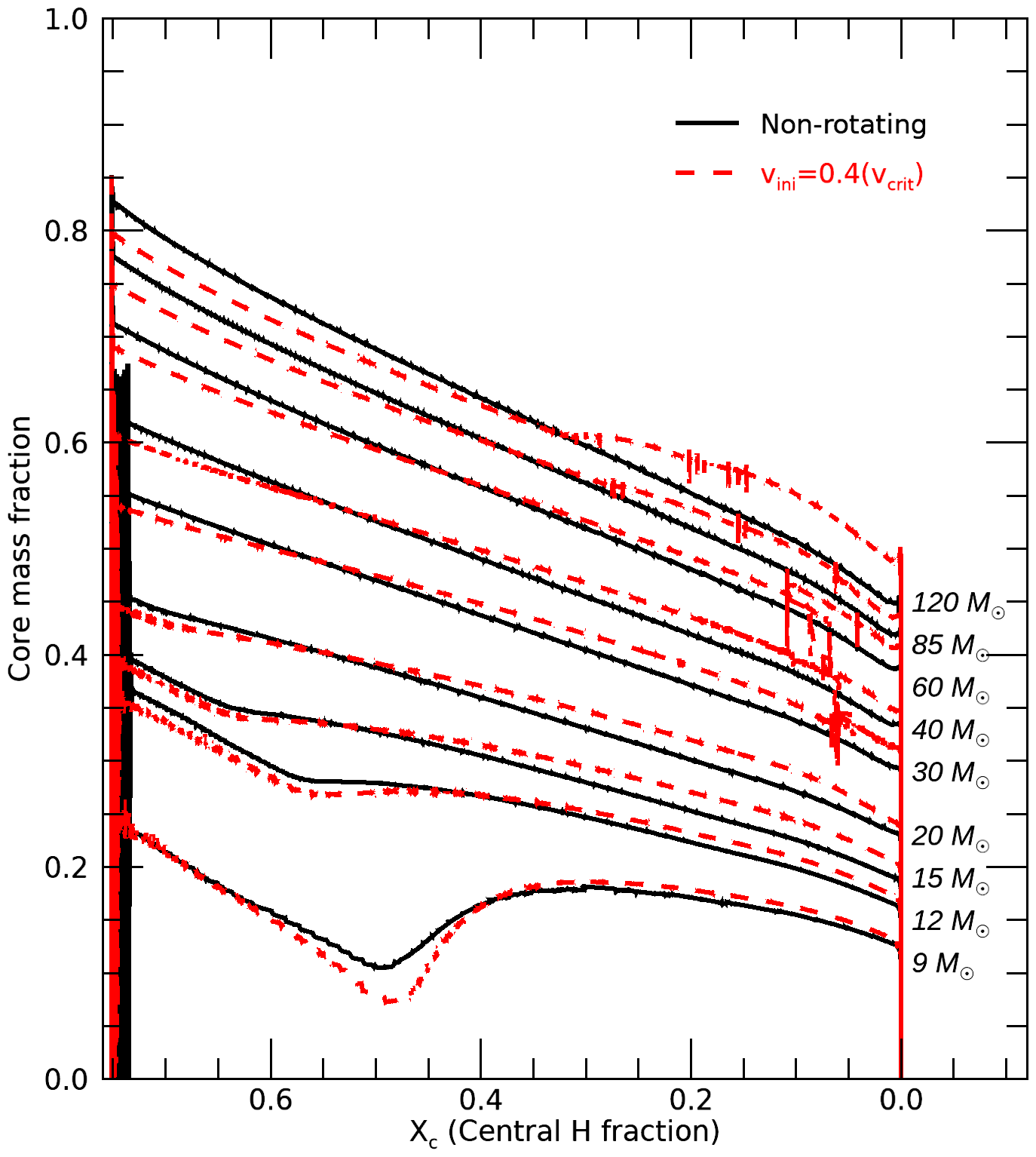}
      \caption{Effect of rotation on the stellar core mass fraction on the MS, for rotating models with $\vini= 0.4\,\vcrit$. The central H mass fraction is used here as a proxy for time spent on the MS. Also indicated in the plot are the initial masses of the models $9 \,\msun \leq \mini \leq 120\,\msun$.
              }
         \label{fig:mcore}
   \end{figure}

Having discussed key features of Pop~III evolution without rotation, we will now look at how rotation affects the evolution of these models. Looking at \Cref{fig:HRD} we see that rotating models begin H-burning with a lower luminosity than non-rotating models. This is because models with rotation begin their evolution with smaller cores, which is evident from \Cref{fig:mcore} (red dashed lines). Due to the centrifugal force in differentially rotating models, the effective gravity of their cores is lower. This leads to a steeper temperature gradient near the stellar centre, and since the convective core size is determined by the temperature profile of the model, their initial core size is lower than models without rotation. There is a general trend for the MS where rotating models become more luminous than non-rotating models, despite starting with lower luminosity as described above. This growth in luminosity is attributed to rotational mixing bringing additional H into the nuclear burning core and increasing its mass. This is seen in \Cref{fig:mcore} where the cores of rotating models do not decrease at the same rate as those of non-rotators. This effect is more pronounced for more massive models since rotational mixing is more efficient at higher masses. For the most massive models of initial masses $85\,\msun$ and $120\,\msun$, the rotating models have a larger core mass fraction than non-rotating models before the end of the MS. This is because the rotating models have reached critical rotation and experience mass loss, which decreases their total mass. An additional effect is that rotation extends the MS lifetime (see \Cref{gridtable1}), which follows from the rotational mixing of extra H into the core. 

The MS evolution is otherwise similar to non-rotating models, with the exception of the jagged MS evolution for very massive models of 85$\,\msun$ and $120\,\msun$.  This is a consequence of evolution near critical rotation which will be discussed in \Cref{sec:criticalrot}. The $60\,\msun$ model experiences increased mass loss due to critical rotation between the MS and the He-burning phase which leads to an increase in surface temperature, while the $85\,\msun$ model experiences mass loss shortly before the end of the MS, and the $120\,\msun$ model has experienced mass loss while on the MS. As a result, for some of the more massive models, there is some change to the surface properties between the end of the MS and the beginning of He-burning. The $60\,\msun$ model is a good example of this where there is an evolution towards higher $\teff$ before He-ignition.  

The He-burning phase itself brings even more variation when rotation is considered. There is no obvious trend, with some models becoming significantly more luminous like the $9\,\msun$ and $20\,\msun$ cases, while others evolve to lower $\teff$, the most obvious example of this among massive models being the $12\,\msun$ model. This indicates that the evolution becomes much more complex when we consider rotation, and there is much to investigate as we examine the interior structure and energy generation of these models. As a result of rotation, the less massive models have lower $\teff$ at the end of He core burning with respect to non-rotating models, while more massive models (85$\,\msun$ and $120\,\msun$) end their He-burning phase with higher $\teff$, likely due to their mass loss at the critical rotation limit, see \Cref{sec:criticalrot}. 

The intermediate mass models are greatly impacted by rotation. They become much more luminous, with the $1.7\,\msun$ model increasing in luminosity by $\sim$ 1 dex and reaching the same luminosity as the rotating $3\,\msun$ model. Rotating models also evolve to much lower $\teff$ than non-rotating ones, favouring red giant formation. This presents interesting possibilities for their evolution, especially if they are part of a binary system as the large radius would favor interaction with a companion.


\begin{figure*}
    \centering
    \begin{subfigure}[t]{0.5\textwidth}
        \centering
        \includegraphics[height=4.5cm]{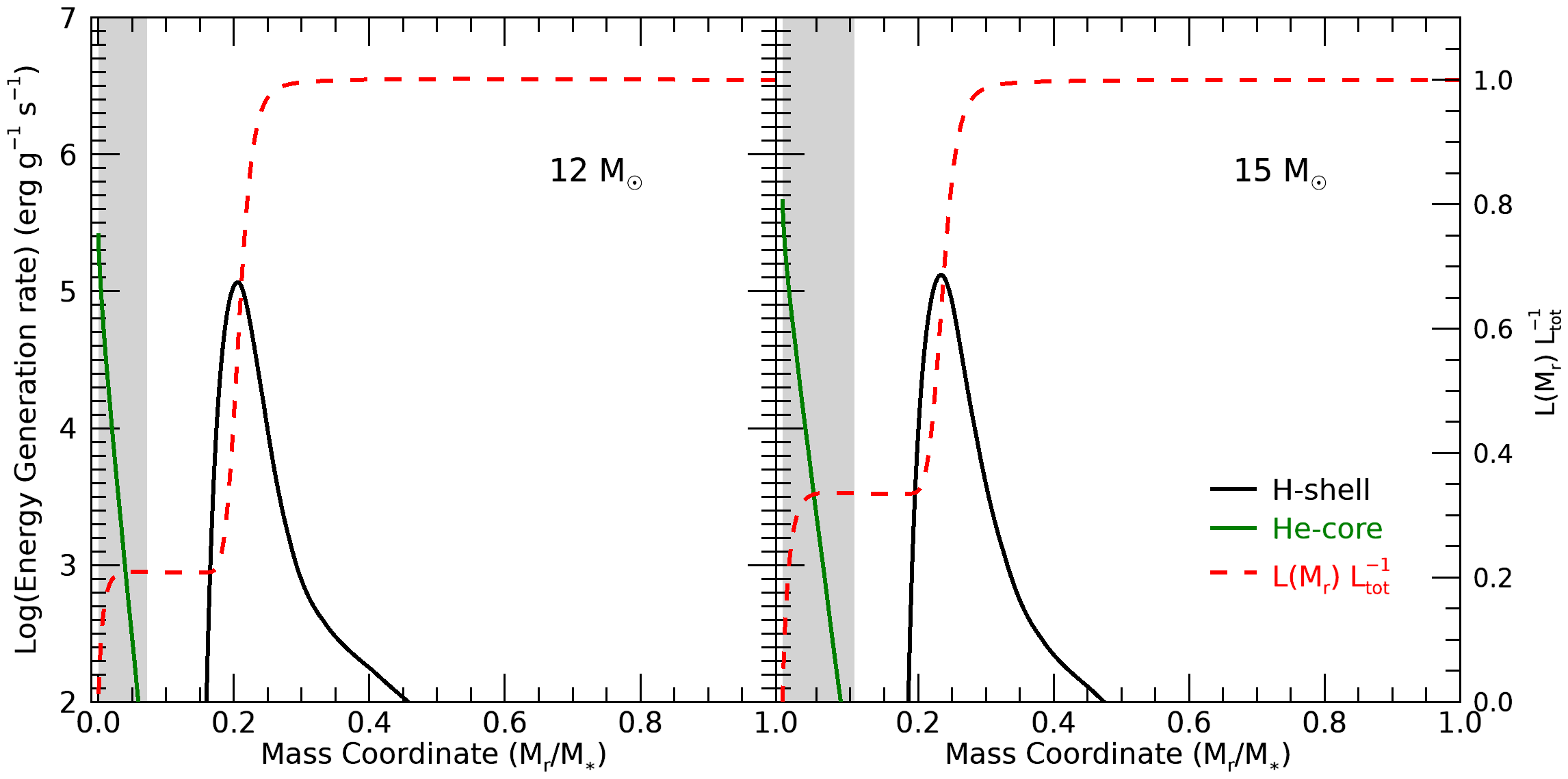}
        \caption{Energy Generation: $Y_{\rm c}$=0.75}
            \label{subfig:Energy_12v15a}
    \end{subfigure}%
    \begin{subfigure}[t]{0.5\textwidth}
        \centering
        \includegraphics[height=4.5cm]{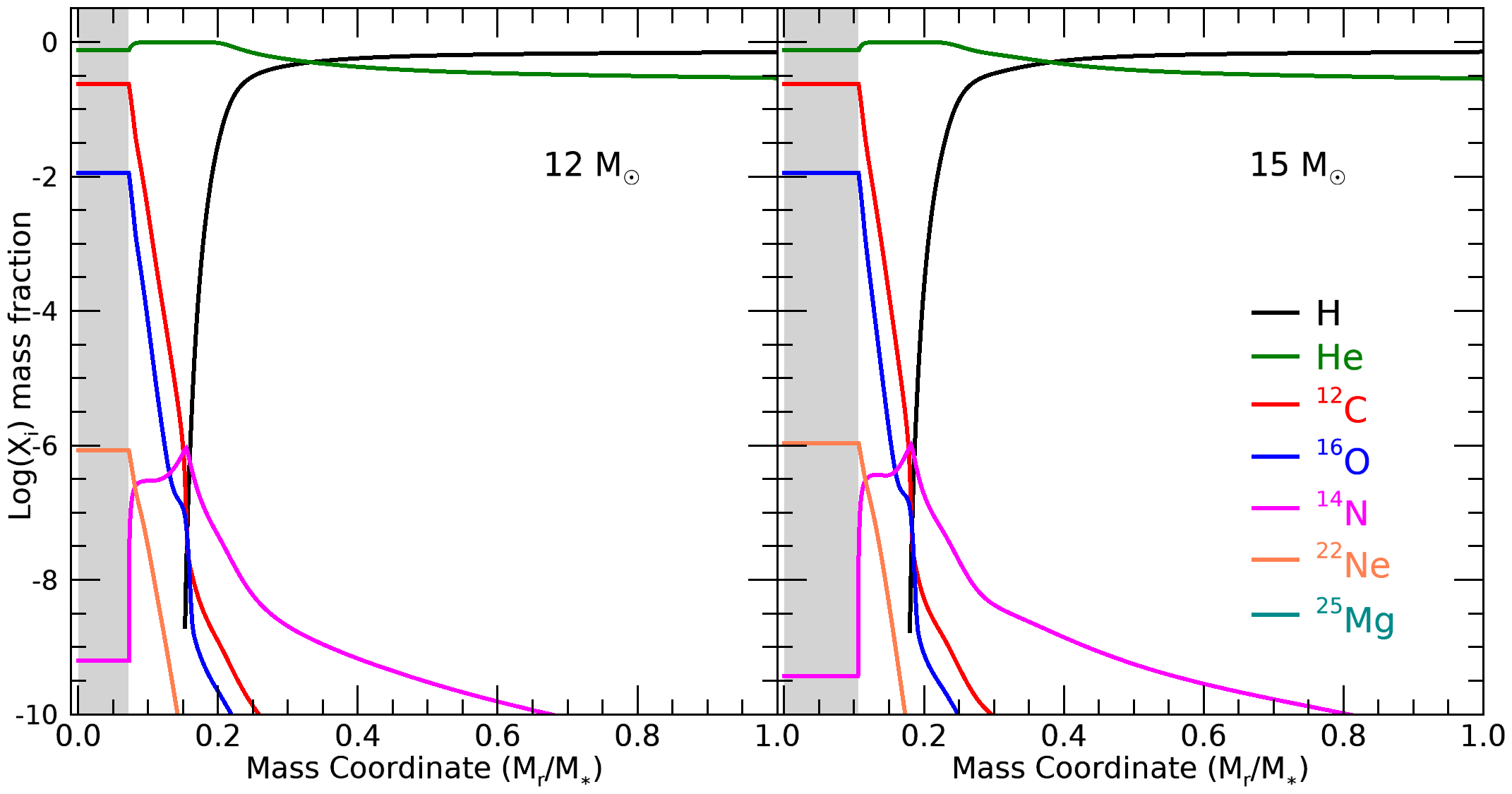}
        \caption{Abundance profile: $Y_{\rm c}$=0.75}
        \label{subfig:Abund_12v15a}
    \end{subfigure}\vspace{0.5cm}
    \begin{subfigure}[t]{0.5\textwidth}
        \centering
        \includegraphics[height=4.5cm]{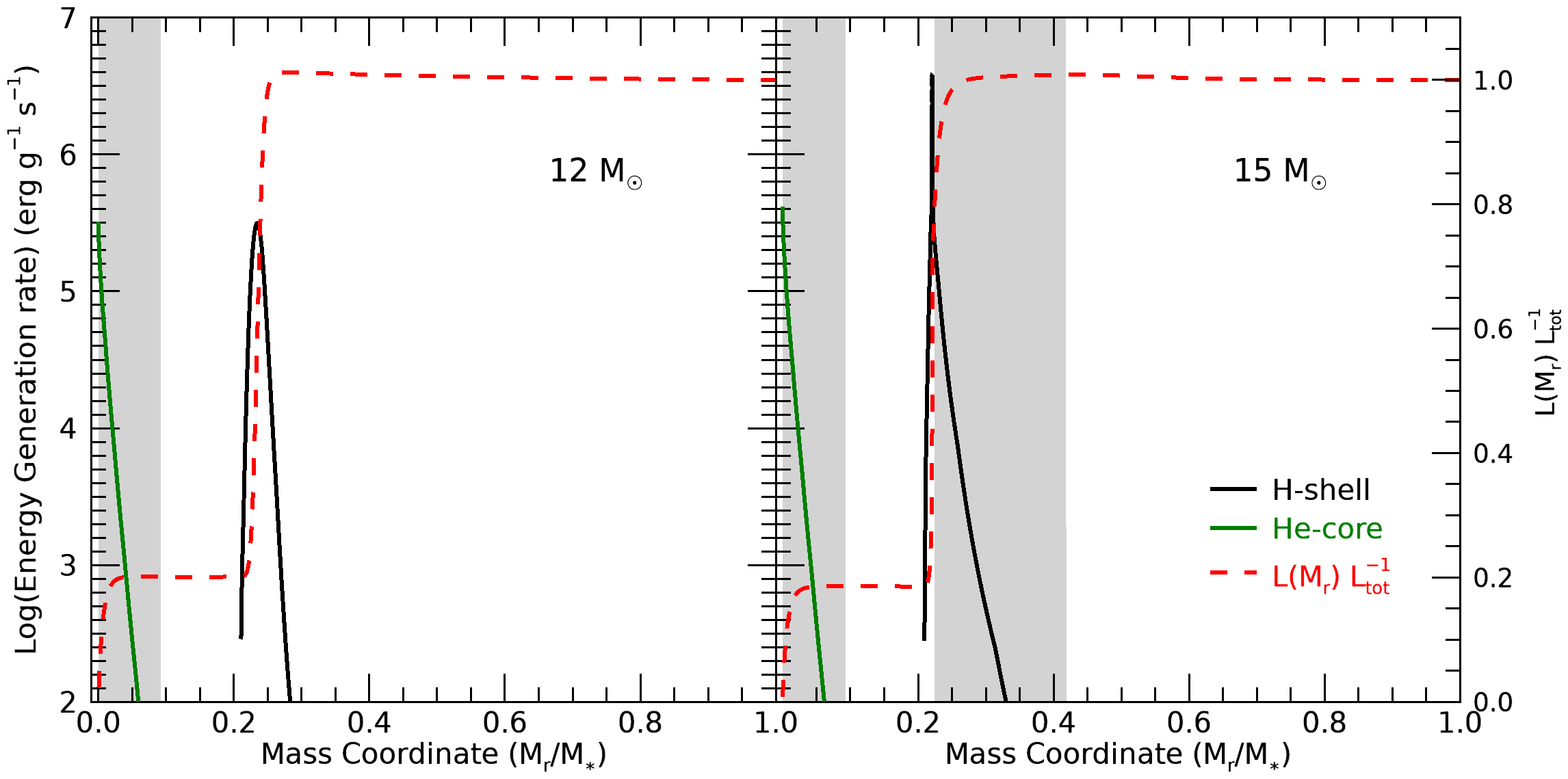}
        \caption{Energy Generation: $Y_{\rm c}$=0.5}
        \label{subfig:Energy_12v15b}
    \end{subfigure}%
    \begin{subfigure}[t]{0.5\textwidth}
        \centering
        \includegraphics[height=4.5cm]{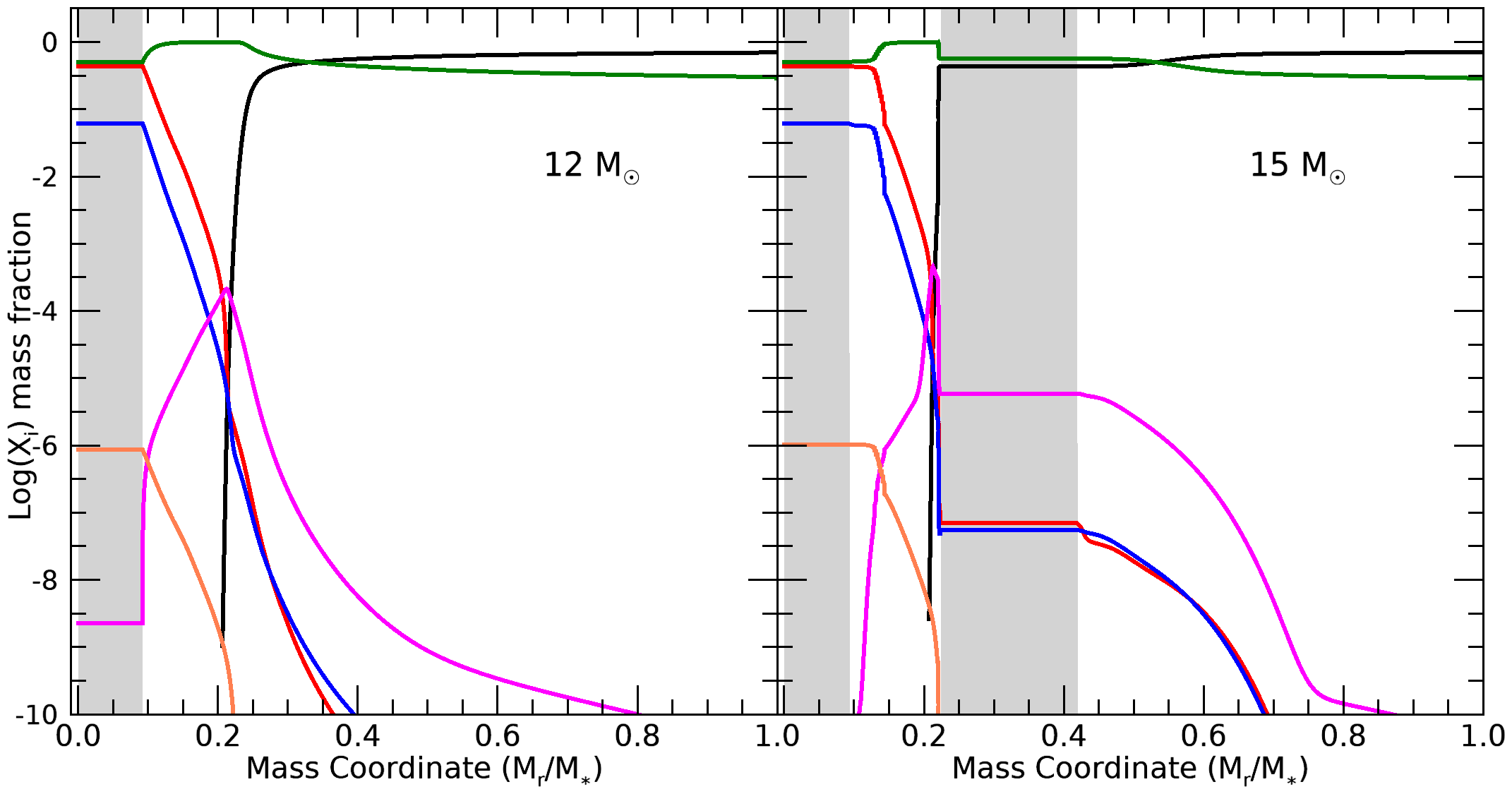}
        \caption{Abundance profile: $Y_{\rm c}$=0.5}
        \label{subfig:Abund_12v15b}
    \end{subfigure}\vspace{0.5cm}
    \begin{subfigure}[t]{0.5\textwidth}
        \centering
        \includegraphics[height=4.5cm]{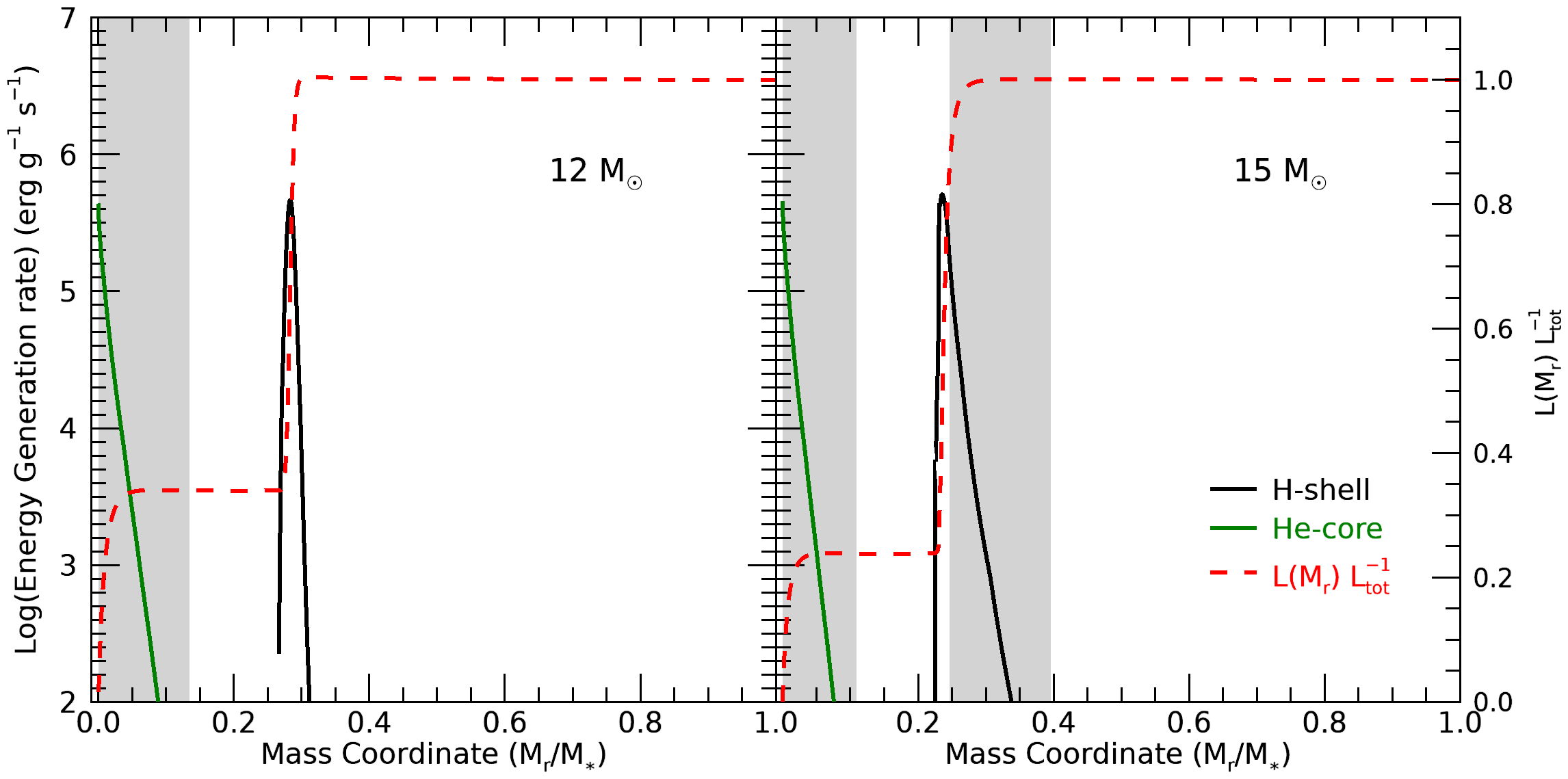}
        \caption{Energy Generation: $Y_{\rm c}$=0.25}
        \label{subfig:Energy_12v15c}
    \end{subfigure}%
    \begin{subfigure}[t]{0.5\textwidth}
        \centering
        \includegraphics[height=4.5cm]{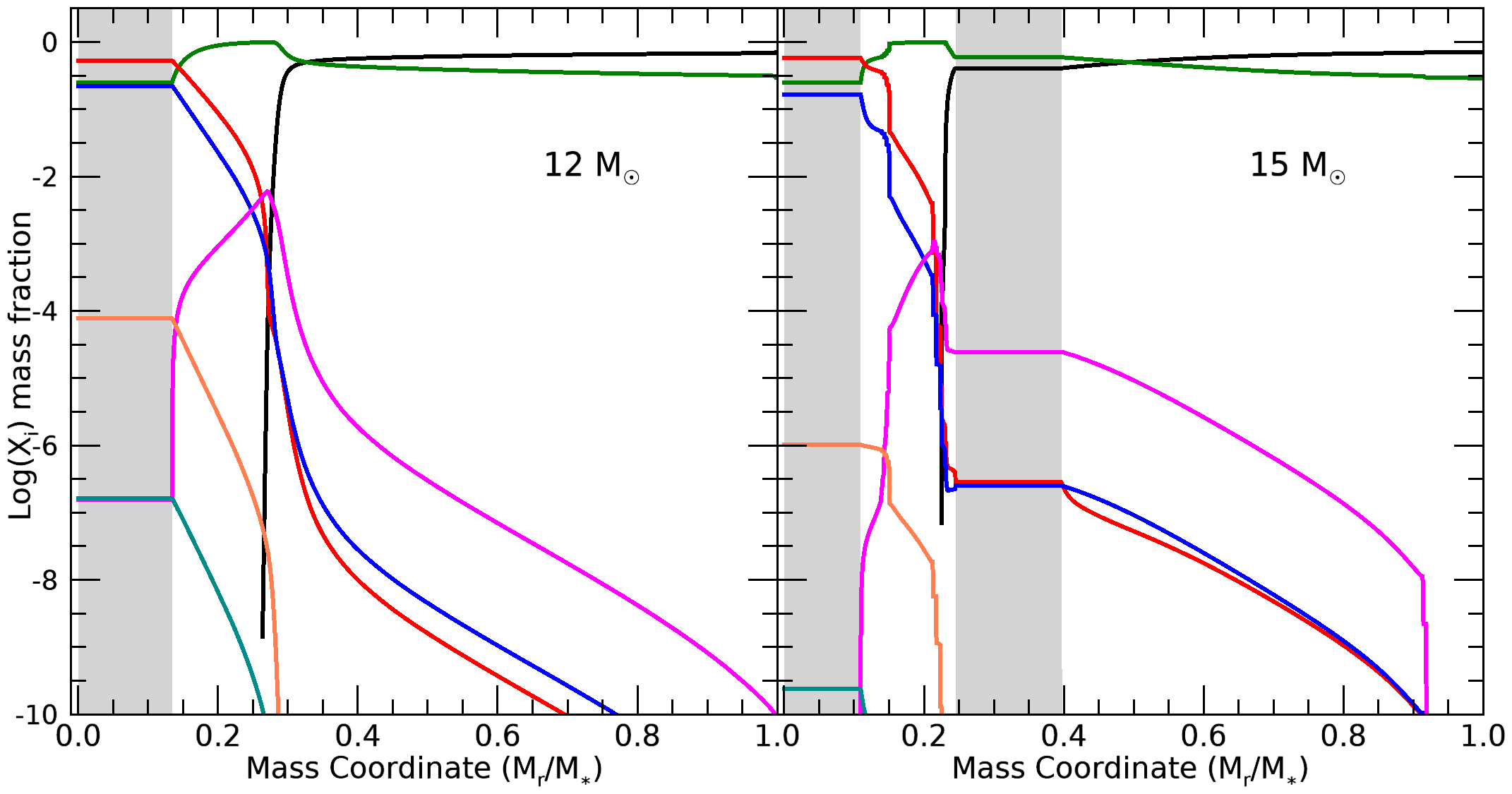}
        \caption{Abundance profile: $Y_{\rm c}$=0.25}
        \label{subfig:Abund_12v15c}
    \end{subfigure}
    \caption{Energy generation rates (left panels) and abundance profiles (right panels) of $12\,\msun$ and $15\,\msun$ rotating models at three separate points of He-burning indicated by their central He fraction, $Y_{\rm c}$. For energy generation rate plots (panels a, c and e), green solid lines indicate energy generation from He-burning and energy generation from H-burning is given by black solid lines. The fraction of luminosity contribution is given by the red dashed lines with values shown in the y-axis on the right-hand-side, for example in the $12\,\msun$ case the energy from the core contributes to 20\% of the total luminosity at $Y_{\rm c}=0.75$. Abundance profiles (panels b, d and f) show chemical abundances throughout the star from centre to surface where species are indicated by the legend (panel b). Convective regions are indicated by the grey shaded areas.}
    \label{fig:Energy_12v15}
    
\end{figure*}


Probably the most significant result from \Cref{fig:HRD} is the variability in the behaviour of rotating models during He-burning. Unlike non-rotators, which show a trend of larger decrease of $\teff$ with increasing mass, rotating models seem to experience a variety of evolutionary behaviour on the HR diagram with changing initial mass. This challenges us to question what drives the evolution along the HR diagram, or more specifically, what evolutionary behaviour during He-burning dominates the evolution of surface properties. What we have found is that the evolution of luminosity and effective temperature during He-burning is moderated by a balance of the relative strength of the He core and the H-burning shell. The dominant effects are that a larger core increases luminosity, and a stronger H shell decreases effective temperature, although convection in the shell affects the structure of the star and subsequently the effective temperature.

\subsection{Internal structure} \label{subsec:internalstruc}

We will focus on the $12\,\msun$ and $15\,\msun$ models first to visualise the complex effects of rotation on internal structure. These models are a good example of the diversity in post-MS surface evolution, with the $12\,\msun$ model experiencing a significant decrease in surface temperature reaching a $\teff$ of almost $10^{4}$ K (right panel of \Cref{fig:HRD}) before the end of He-burning, while the $15\,\msun$ shows more variance in luminosity but has a higher $\teff$ of roughly $10^{4.4}$ K ($\sim$ 25120 K) at the end of He-burning. By looking at the interior structure of our rotating models it is clear that the nature of the H shell plays a dominant role in determining the model's structure and behaviour during the He-burning phase. 

We show in \Cref{fig:Energy_12v15} the interior structure of the $12\,\msun$ and $15\,\msun$ rotating models with $\vini=0.4\,\vcrit$ at three key points of He-burning evolution. We use the central He fraction ($Y_{\rm c}$) as a reference point for evolutionary stage, where $Y_{\rm c}$=1 would indicate the start, and $Y_{\rm c}$=0 the end, of He-burning. At $Y_{\rm c}$=0.75 it can be seen that these models have a very similar structure both in abundance and in nature of the H shell (\Cref{subfig:Energy_12v15a,subfig:Abund_12v15a}). The only difference between them is the size of the He-burning core where the $15\,\msun$ model's core is a larger fraction of the total mass. This is an important difference because more He-burning products are then transported to the H shell through rotational mixing. Indeed this mixing of He-burning products, such as $^{12}\mathrm{C}$ and $^{16}\mathrm{O}$, towards the H shell can be seen in the abundance profile (\Cref{subfig:Abund_12v15a}). 

Given that these are zero-metallicity models, the H shell relies on p-p chain reactions for nuclear burning until $^{12}\mathrm{C}$ and $^{16}\mathrm{O}$ reach these regions through rotational mixing. The introduction of these heavier elements triggers the CNO cycle, which significantly boosts energy generation in the H shell \citep{Ekstrom2008EffectsStars}. While the $12\,\msun$ abundance profile shows that these elements mix outwards from the core they do not trigger a strong CNO boost (left panels of \Cref{subfig:Energy_12v15b,subfig:Abund_12v15b}), indicating that insufficient He-burning products reached the H shell for the CNO cycle to dominate H-burning. The H-burning shell in the $12\,\msun$ model therefore remains dominated by p-p chain reactions and radiative. As this model evolves, the He core continues to grow while the H shell moves outwards (left panels of \Cref{subfig:Energy_12v15c,subfig:Abund_12v15c}). This outward evolution of the H shell makes sense, because as it produces Helium and depletes Hydrogen in one layer of the star, it must move closer to the surface to source layers richer in Hydrogen and continue burning. The evolution of the temperature profile may also play a role in how the H shell moves outwards as regions closer to the stellar surface become hot enough for H-burning. 

It can be seen from the energy generation profiles for this $12\,\msun$ model, that the H shell is a significant source of luminosity for the star. It contributes to about 80\% of the total luminosity at $Y_{\rm c}$=0.75 (left panel of \Cref{subfig:Energy_12v15a}),
and 65\% at $Y_{\rm c}$=0.25 (left panel of \Cref{subfig:Energy_12v15c}). Consequently, the changes to the H shell strongly impact the structure of the star, and are related to an increase in the stellar radius as the H shell evolves outwards. This explains the large decrease in $\teff$ that we observe in \Cref{fig:HRD} for the $12\,\msun$ model. The H shell dominates the total energy contribution and, therefore, the stellar structure, forcing the star to adopt a larger radius in order to maintain hydrostatic equilibrium. This behaviour is evident from our models given that the radius begins increasing at $Y_{\rm c}$=0.6, just as the H shell begins moving outwards through the stellar envelope. We note here that determining the dominant factors for the evolution in the HR diagram is complex. According to Farrell et al. (in prep.), there are four main factors that drive the evolution to lower effective temperatures during central He-burning. These are: an increase in He abundance in the H-burning shell, an increase in the core mass ratio in the regime of $M_{\rm core}/M_{\rm total} > 0.6$, an increase in the CNO abundance in the H-burning shell, and a decrease in the central He abundance ($Y_{\rm c}$) during the latter half of core He-burning.

In contrast to the $12\,\msun$ model, the $15\,\msun$ model can more easily mix these elements into its H shell because of its larger core (\Cref{subfig:Energy_12v15a,subfig:Abund_12v15a}) which results in a strong CNO boost at $Y_{\rm c}$=0.52. \Cref{subfig:Energy_12v15b,subfig:Abund_12v15b} illustrate the consequences of this CNO boost shortly after its occurrence. While the temperature dependence of p-p chain reactions is roughly $\epsilon_{pp} \propto T^4$, the CNO cycle has a much higher temperature dependence of $\epsilon_{CNO} \propto T^{20}$. Therefore, the CNO cycle steepens the temperature gradient at the boundaries of the H shell which triggers convection. The CNO boost is named as such because of the effect that it has on energy generation, if we compare the $15\,\msun$ model in \Cref{subfig:Energy_12v15a,subfig:Energy_12v15b} we can see this effect through the increased luminosity contribution of the H shell at $Y_{\rm c}$=0.5. Earlier in the He-burning phase the H shell of the $15\,\msun$ model contributed to approximately 70\% of the total luminosity (right panel \Cref{subfig:Energy_12v15a}), while after the CNO boost its contribution is more than 80\% (right panel \Cref{subfig:Energy_12v15b}). This boost in energy of the shell causes the He-burning core to retract, an effect which is also seen in \citet{Ekstrom2008EffectsStars}. This explains the decrease in luminosity observed in the right panel of \Cref{fig:HRD}, where a dip in $\logl$ is seen at approximately $\logteff$=4.6. Now that the H shell is convective (right panels \Cref{subfig:Energy_12v15b,subfig:Abund_12v15b}) it can maintain strong H-burning in this region by replenishing its fuel through convective mixing (right panels \Cref{subfig:Energy_12v15c,subfig:Abund_12v15c}). It also maintains the current structure of the star, preventing the radius from increasing at the same rate as the $12\,\msun$ model, which explains why the $15\,\msun$ model does not reach values of $\teff$ as low as that of the $12\,\msun$ model.

Comparing the H-profile of the $12\,\msun$ model and the $15\,\msun$ model in the abundance profile at $Y_{\rm c}$=0.25 (\Cref{subfig:Abund_12v15c}) shows that the lower mass model has a higher H abundance in the envelope. Given that the opacity of these models is dominated by electron scattering the higher H abundance infers a higher opacity, so it makes sense that the $12\,\msun$ model reaches the redder part of HR diagram in \Cref{fig:HRD}. However, this does not drive the increasing radius. The energy provided by the outwards moving H shell to regions closer to the stellar surface drives the stellar expansion. The nature of the H shell is, therefore, responsible for the strong variation in evolution along the HR diagram during He-burning, as this work confirms. Through investigating these two models we have also shown how sensitive this H-burning shell is to products of He-burning that diffuse out from the core through rotational mixing.  

   \begin{figure}
   \centering
   \includegraphics[width=\linewidth]{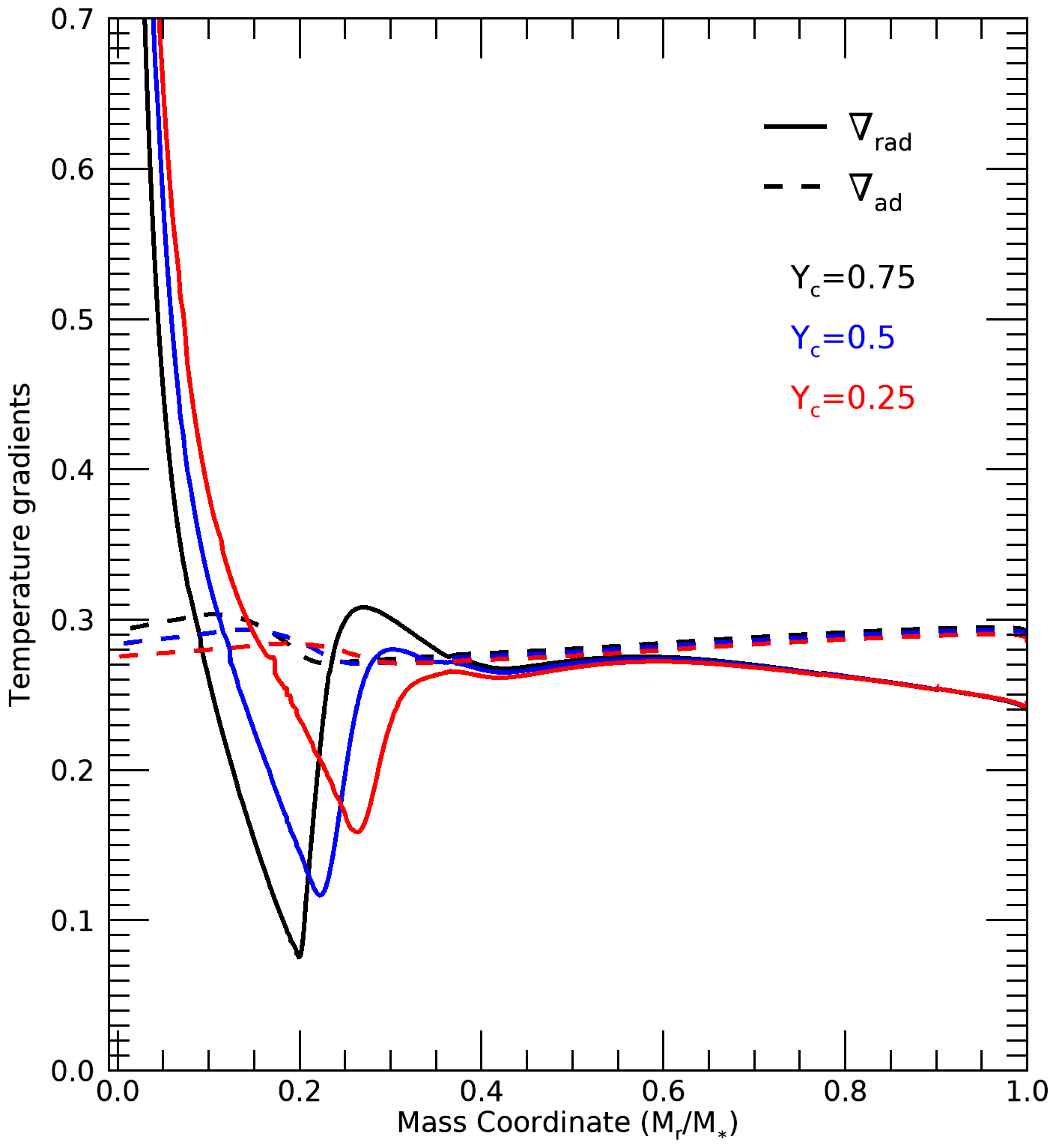}
      \caption{Evolution of the radiative and adiabatic gradients through He-burning for the $20\,\msun$ model rotating at an initial velocity $\vini=0.4 \, \vcrit$. Central He fraction ($Y_{\rm c}$) is indicated by the legend. Convection regions appear where $\nabla_{\rm rad} > \nabla_{\rm ad}$.   
              }
         \label{fig:tempgrads20}
   \end{figure}
%

Now that we better understand the complexities of how rotation affects stellar structure in these models, let us compare the behaviour of the 12 and $15\,\msun$ models to those of lower and higher \mini. In the $9\,\msun$ model, the H shell is not significantly stronger than the He core and so we do not see a large decrease in $\teff$ (\Cref{fig:HRD}). We do observe a considerable increase in luminosity however, which is indicative of the growing He core aided by rotational mixing.

The 20, 30, and $40\,\msun$ rotating models behave similarly to each other in the HR diagram. They each experience a substantial decrease in $\teff$ and a boost to their luminosity during He-burning. This increase in luminosity results from the growth of the He core, which also triggers a reduction in the size of the convective H shell. In the $20\,\msun$ model, the H shell actually becomes radiative again. The growth of the He-burning core affects the temperature profile of the star which in turn changes the radiative temperature gradient ($\nabla_{\rm rad}$). This is shown in \Cref{fig:tempgrads20} where $\nabla_{\rm rad}$ and the adiabatic temperature gradient ($\nabla_{\rm ad}$) are plotted for three stages of the He-burning phase ($Y_{\rm c}=0.25,0.5,0.75$.) The figure shows that as the core grows in size it flattens the radiative gradient profile. Since we only have convection where $\nabla_{\rm rad} > \nabla_{\rm ad}$, the convective region reduces in size until the H shell becomes radiative. Models from \citet{Farrell2020_snapshot} support this conclusion in showing that for higher core mass ratios, the value of $\nabla_{\rm rad}$ is lower which tends to disfavour convection. Similarly to the $12\,\msun$ model, the now radiative H shell moves further towards the stellar surface driving expansion and ends up in a redder part of the HR diagram (\Cref{fig:HRD}) than its non-rotating counterpart. 

In summary, we find that the evolution of surface properties during He-burning is moderated by a balance of the relative strength of the He~core and the H-burning shell, and how this impacts the temperature profile. In some cases the H shell affects the size and strength of the He~core, for instance when the CNO boost causes the core to retract. In other cases, the He core affects the size and strength of the H~shell, for instance when the growth of the He~core flattens the temperature profile and removes convection from the H~shell. These effects are particularly important for fast rotating models where rotational mixing leads to increased energy production and changes to the chemical profile, affecting metal enrichment.

   \begin{figure}
   \centering
   \includegraphics[width=\linewidth]{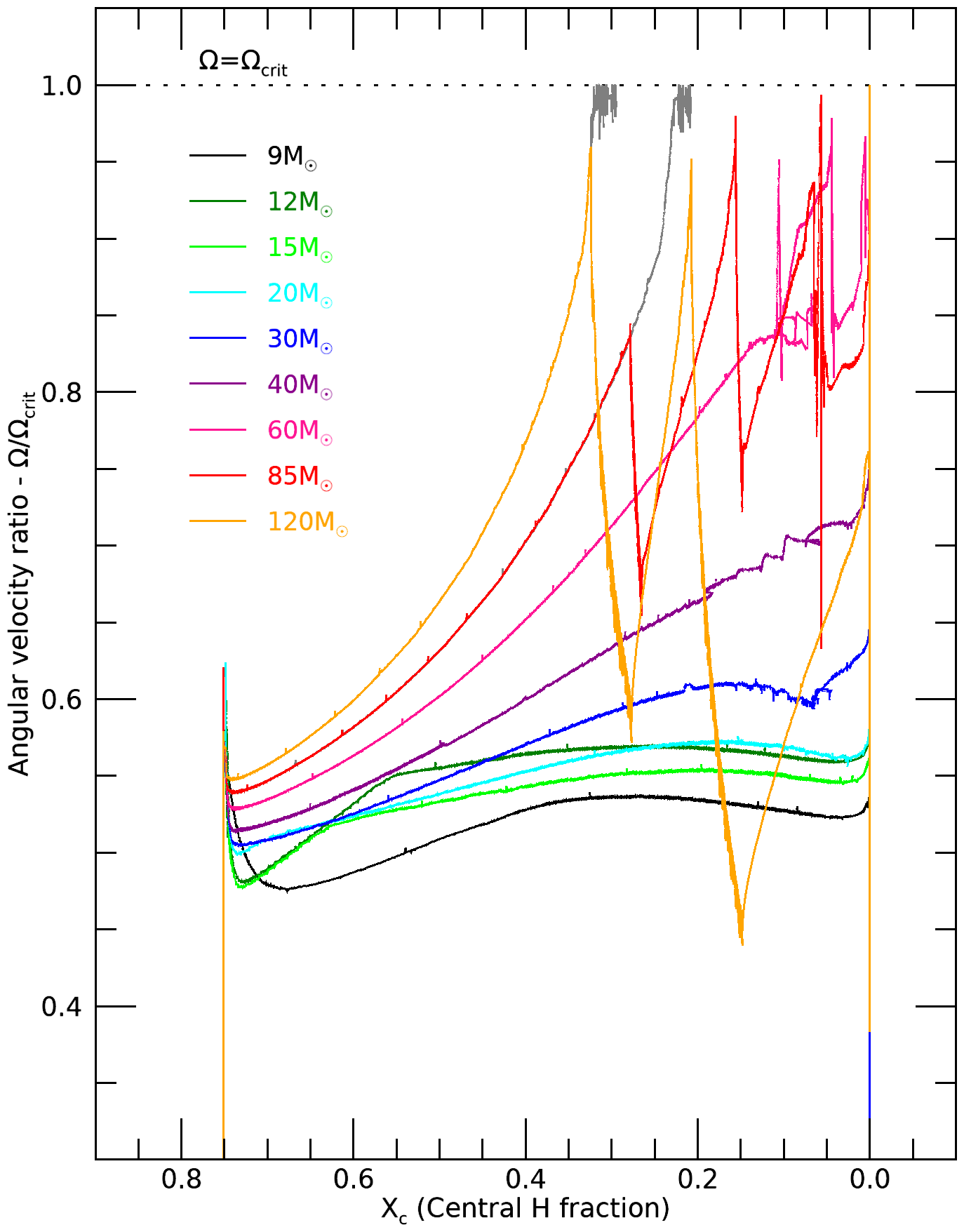}
      \caption[Angular velocity ratio evolution during the MS]{Evolution of the angular velocity ratio, $\Omega/\Omega_{\rm crit}$, during the MS for models with $\vini=0.4\,\vcrit$, initial masses are indicated in the legend. The dotted line indicates where models have reached critical velocity, the grey lines show the $85\,\msun$ and $120\,\msun$ models reaching critical when we rely on the mechanical mass loss implementation in the code. Their corresponding coloured lines show where spherical mass loss is added.  
              }
         \label{fig:avel40}
   \end{figure}
%
   \begin{figure}
   \centering
   \includegraphics[width=\linewidth]{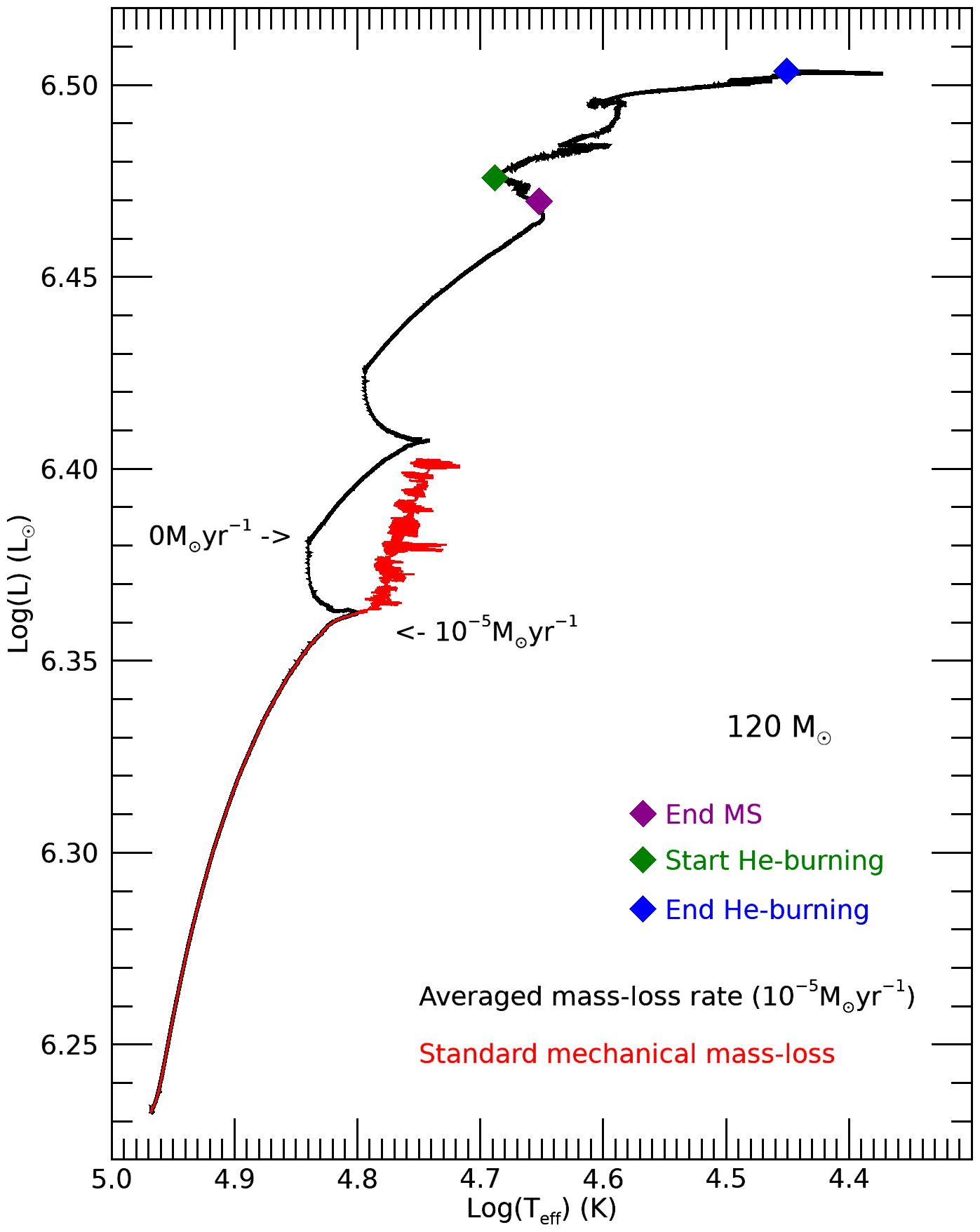}
      \caption{Evolution of the $120\,\msun$ model until the end of He-burning as indicated in the legend. The black evolutionary track corresponds to the model where a mass loss rate of $10^{-5}\,\msunyr$ is switched on as the model approaches critical velocity (see arrows). The red evolutionary track corresponds to our model with mechanical mass loss. 
              }
         \label{fig:120HR}
   \end{figure}

%

\subsection{Final fates and proximity to pair instability}
\label{subsec:finalfates}

Models from this grid have been used to investigate the possibility of forming primordial black holes within the so-called pair instability mass gap. This is discussed in \citet{Farrell2020_GW190521} as a possible explanation for the black hole masses detected in the recent GW190521 merger event \citep{ligo20report,ligo20implications}. Our zero metallicity models are promising candidates for the black hole mass required given their negligible mass loss and compact structure. Some models, e.g. the non-rotating 85$\,\msun$ model, achieve lower CO core masses through H-He shell interactions, which may help them avoid the pair instability regime. Other works have also suggested Pop~III stars as potential progenitors for the GW190521 merger \citep{Kinugawa2020,Liu2020,Safarzadeh2020,Tanikawa2020}. \citet{Umeda2020} found that even if the CO core mass reaches the pulsational pair instability limit, these stars could remain mostly intact if their binding energy is high enough.

Since Pop~III stars are more compact than higher metallicity stars, this is further evidence that they may raise the lower limit of the pair instability mass gap. This is in contrast to \citet{Chatzopoulos2012} which suggested that rotational mixing would increase the core size of Pop~III stars leading them to encounter the pair instability at lower initial masses than higher metallicity stars. The CO core (M$_{\mathrm{CO}}$) and He core (M$_{\mathrm{He}}$) masses for our models are given in \Cref{gridtable2}. We find that rotational mixing does not lead to a general increase in core sizes at late evolutionary stages, and in fact most models show lower CO core mass with rotation. Rotational mixing does increase the core size during the MS, however, this is not necessarily true for the post-MS stages. This is largely because from He-burning onwards rotational mixing strengthens the H-burning shell which tends to suppress the growth of the He core. The differences in behavior of our models compared to earlier work could be due to the different assumptions about convection and rotational mixing. Given their impacts on the final core mass of Pop~III stars, further work is warranted on a detailed exploration of the physics of mixing in Pop~III stars.


\section{Critical Rotation and Mass Loss}\label{sec:criticalrot}
  
Given the lack of radiative mass loss in Pop~III models, there is no mechanism for removing angular momentum from the surface of these stars. While meridional currents are weak in zero-metallicity stars due to their higher density \citep{Ekstrom2008EffectsStars}, angular momentum is still transported outwards from the core, and without mass loss this angular momentum builds at the surface. As a consequence, several of our models spin up during the MS. This can be seen in \Cref{fig:avel40}, where the evolution of the angular velocity on the MS is plotted. The angular velocity is plotted as a fraction of critical, that is, the velocity at which the outer layers of the star become unbound. The dotted horizontal line indicates this point clearly, and allows us to observe how our models evolve towards this limit during H-burning. As can be seen from the figure, models $>\!20\,\msun$ spin up on the MS with more massive ($\geq\!60\,\msun$) models reaching critical. The grey lines show the $85\,\msun$ and $120\,\msun$ models reaching critical rotation when we rely on the mechanical mass loss implementation in the code. The red and orange lines correspond to the $85\,\msun$ and $120\,\msun$ models where our averaged mass loss rate was switched on when the star approaches critical rotation. To explain this difference in mass loss treatment we will look specifically at the $120\,\msun$ example.   

When a model reaches critical velocity the outer layers of the star become unbound and it can be expected that the star will lose a significant amount of mass at this point, enough to lose sufficient angular momentum to fall below critical velocity again. This is the case for the $120\,\msun$ model with $\vini=0.4\,\vcrit$ which reaches critical velocity mid-way through the MS. Its evolution along the HR diagram is shown in \Cref{fig:120HR}. The red line shows the evolution of the star with only the mechanical mass loss implementation, described in \Cref{sec:ingredients}, to simulate mass loss at critical rotation, and corresponding to the grey lines in \Cref{fig:avel40}. With this mechanical mass loss prescription the model remains at critical rotation, giving rise to the unstable region in \Cref{fig:120HR} at luminosities 6.36 $\leq\logl\leq$ 6.4, and the evolution along the $\Omega=\Omega_{\rm crit}$ line in \Cref{fig:avel40}. To resolve this instability we impose an averaged mass loss rate of $10^{-5} \,\msunyr$ just before the model reaches critical velocity. This treatment is shown by the black line in the HR diagram (\Cref{fig:120HR}) and the orange line in \Cref{fig:avel40}. 

As the star spins up, its outer layers expand lowering the surface temperature and evolving the star to the right of the HR diagram. Upon reaching critical, our higher mass loss rate was employed in order to allow the model to shed the unbound mass from the outer envelope (see arrow in \Cref{fig:120HR}). This increases the surface temperature as deeper layers of the envelope are revealed. Once enough mass is lost to bring the rotational velocity below critical, the surface temperature stabilises and the increased mass loss rate raises the luminosity. The mass loss rate was then switched off again (see arrow in \Cref{fig:120HR}) so as to allow the model to resume its MS evolution. However, as is evident from \Cref{fig:avel40,fig:120HR}, such a mass loss event can reoccur if the model can spin up to critical again. This shows that angular momentum transport is efficient enough to replenish the angular momentum lost at the surface. As a result, this behaviour may occur multiple times before H core depletion, depending on the initial mass and rotation of the star. During the first period of mass loss we see the velocity decrease as angular momentum is lost through the sharp dip in \Cref{fig:avel40} at 0.32 $\geq X_c \geq$ 0.28. We then see how quickly the model spins up again when mass loss is returned to zero at $\Omega/\Omega_{\rm crit}$=0.59. This illustrates how difficult it is for these massive Pop~III models to evolve away from critical. The total mass lost by the $120\,\msun$ model is $3.5\,\msun$, having spent $\sim\! 8\%$ of its MS lifetime losing mass.

We use a similar mass loss treatment for the 60 and $85\,\msun$ models before they reach critical rotation (pink and red lines in \Cref{fig:avel40}, respectively.). These models experience shorter periods of mass loss since they evolve away from critical more easily. The $85\,\msun$ model loses roughly $1\,\msun$ during these mass loss events, while $0.3\,\msun$ is lost by the $60\,\msun$ model (see \Cref{gridtable1}). The mass loss of the 60, 85 and $120\,\msun$ models also explains the increase in the core mass fraction in \Cref{fig:mcore} towards the end of the MS, when these models reach critical. During this time, the total mass of the star decreases, which leads to an increase in the core mass fraction. This behaviour may have significant impacts for the final fates of these models given the effect of mass loss during the MS on the core mass. Of course, the amount of mass lost by these models is a direct consequence of the assumed value of \mdot\ in our models, and further study is needed to investigate the behavior of mass loss in fast-rotating Pop~III stars.

   \begin{figure}
   \centering
   \includegraphics[width=\linewidth]{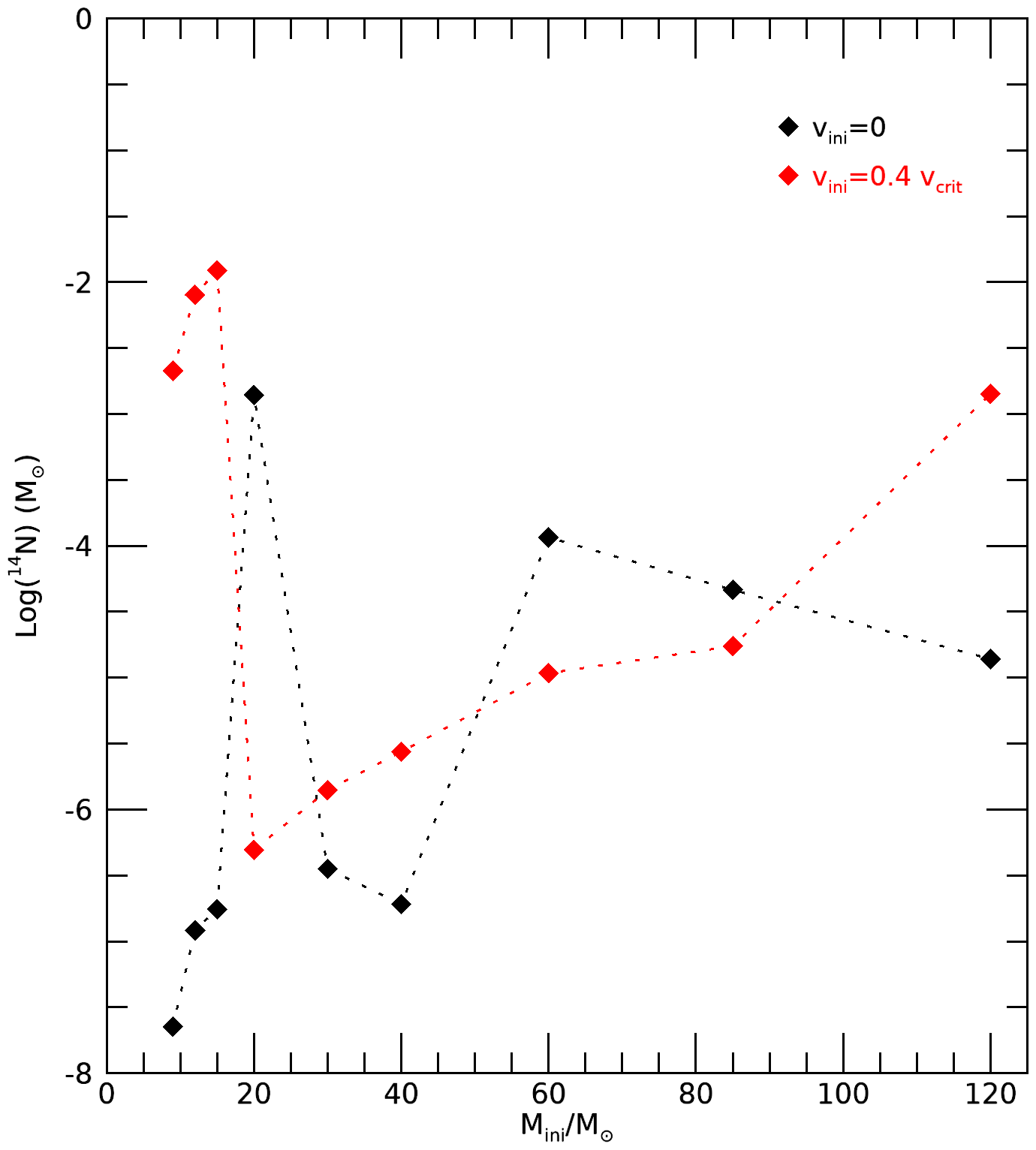}
      \caption{Total mass of $\nitrgn$ above the gravitational remnant mass at final stage in evolution for each model, see \Cref{gridtable2}, with initial mass in units of solar mass, models with rotation are shown in red.
              }
         \label{fig:enrichment}
   \end{figure}
%

\section{Metal Enrichment and Yields}\label{sec:enrichment}

One of the most important aspects of Pop~III evolution is their metal enrichment. As outlined in \Cref{sec:intro}, enrichment of Pop~III stars has been investigated through studying EMP stars which are believed to be direct descendents of zero-metallicity stars and therefore can constrain their metal enrichment \citep{Choplin2018paper,Hartwig2018fingerprintpopIII}. Here we focus on the evolution of Pop~III stars and how their enrichment is affected by initial mass and rotation. By connecting this with work being done on second generation stars we can get a better picture of how the first stars would have evolved and produced the first heavy elements in the universe. 

\Cref{gridtable2} shows the final amounts of \nitrgn, $^{12}\mathrm{C}$, and $^{16}\mathrm{O}$ produced, noting that the evolutionary stage reached varies for each model. These chemical yields are calculated for mass coordinates above the gravitational remnant mass, which represents the estimated mass of the remaining core following a SN explosion. It is computed based on the CO core mass \citep{Maeder92}. The mass of the CO and He cores, shown in \Cref{gridtable2}, are determined using the method by \citet{Heger2000}, where the mass coordinate where H falls below 10$^{-3}$ defines the He core, and similarly where He falls below 10$^{-3}$ defines the CO core. For models where central He is not yet depleted enough for this definition ($\mini=30\,\msun$ with $\vini=0$ and $\mini \geq 40\,\msun$ with $\vini=0.4 \, \vcrit$) we instead take the mass coordinate where 75\% of He has been burned. We note that CO core mass is highly dependent on the treatment of convection \citep{Kaiser2020}, which as we discuss in \Cref{sec:ingredients}, is possibly underestimated here for consistency with Papers I, II and III. As shown in \Cref{gridtable2}, our models are all at various stages of post-MS evolution so this must be accounted for in establishing trends in metal enrichment. While the amount of $^{12}\mathrm{C}$ and $^{16}\mathrm{O}$ produced changes through late burning phases (C-burning and O-burning) the amount of $\nitrgn$ produced remains largely constant, unless it is consumed by a He-burning region as we will discuss. It is therefore an ideal candidate for our study of metal enrichment. 

\Cref{fig:enrichment} shows the $\nitrgn$ produced by non-rotating and fast-rotating ($\vini=0.4 \, \vcrit$) models. It is important to note that the $\nitrgn$ abundance plotted here is that of the final model, which corresponds to a different evolutionary phase reached for each simulation. This $\nitrgn$ abundance is the total mass of $\nitrgn$ above the gravitational remnant mass at the final evolutionary stage for each model. Even without all of the models reaching the pre-SN stages, we may still expect to see more $\nitrgn$ produced with rotation. However, we find that rotating models do not always have higher metal enrichment. Some non-rotating models actually show more enrichment than their corresponding rotating model of the same initial mass, such as the $20\,\msun$ model. The high $\nitrgn$ enrichment in non-rotating models of 20-$30\,\msun$ has also been found in \citet{Chieffi2004} and \citet{Ekstrom2008EffectsStars}. It is therefore not as straightforward as more rotational mixing allows more enrichment, a strong CNO boost and subsequent $\nitrgn$ enrichment can arise from multiple evolutionary behaviours. Let us first examine how enrichment of rotating models varies with initial mass.

   \begin{figure}
   \centering
   \includegraphics[width=\linewidth]{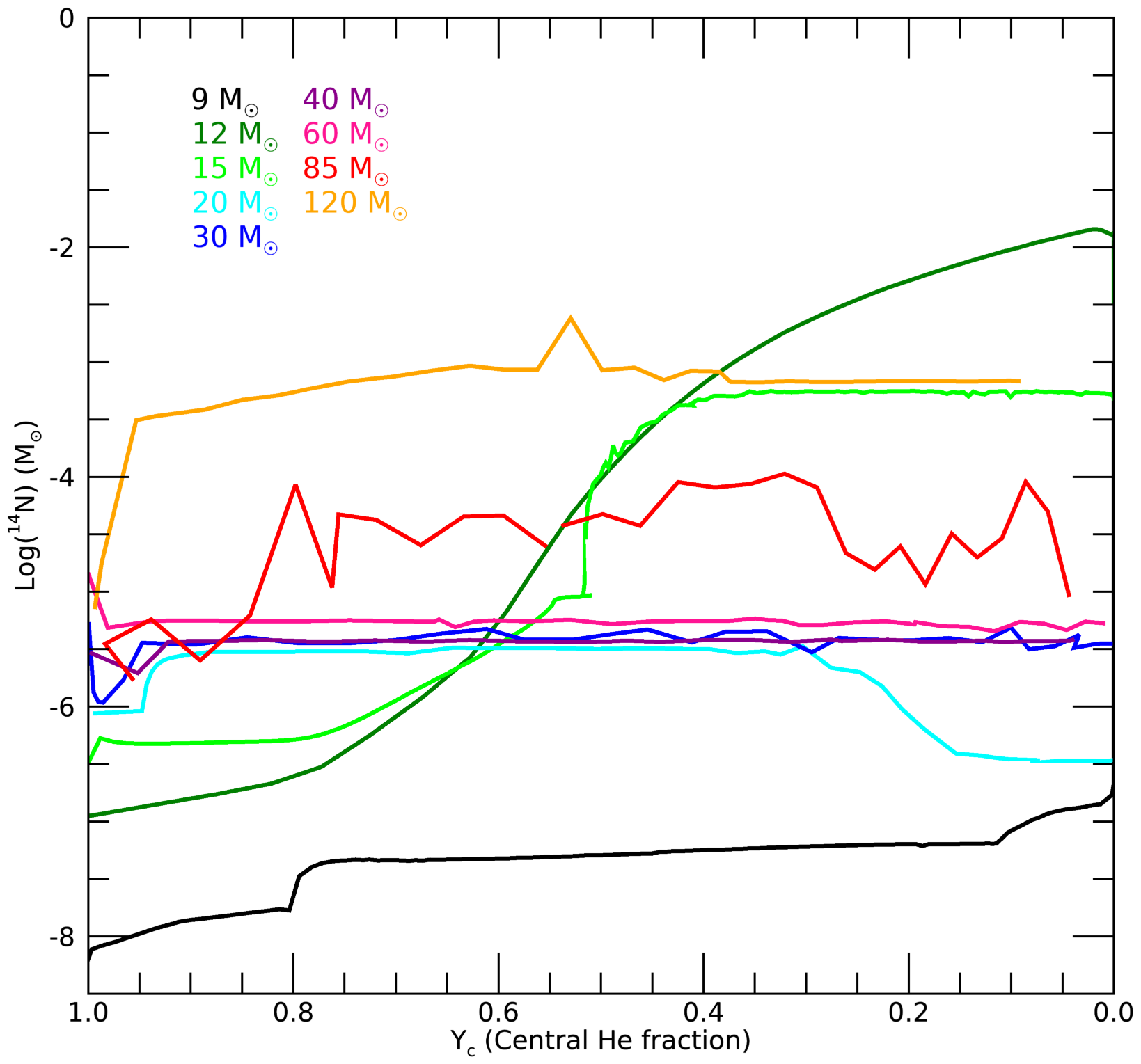}
      \caption{Evolution of $\nitrgn$ abundance during He-burning for models with initial velocity $\vini$=0.4$\vcrit$ and initial masses indicated by legend. Colours are the same as \Cref{fig:avel40}.
              }
         \label{fig:Nevol_rot}
   \end{figure}
%

\subsection{Rotating models}
For the less massive models, $\mini=9,12,15\,\msun$, we do see a trend with rotation, where rotating models show significantly more enrichment. To investigate this behavior, we show in \Cref{fig:Nevol_rot} the evolution of the $\nitrgn$ abundance during He-burning, when we expect most of $\nitrgn$ production to occur. Again the $\nitrgn$ abundance is the total mass of $\nitrgn$ above the gravitational remnant at each stage in the evolution, represented here by the central He fraction. 

We see that $\nitrgn$ abundance remains low for the $9\,\msun$ model throughout He-burning, and the high $\nitrgn$ content seen in \Cref{fig:enrichment} actually results from an interaction between the H and He shells following He-burning. This highlights the importance of understanding how metal enrichment in these stars occurs, since significant $\nitrgn$ production occurs whenever He-burning products interact with a H-burning region. Similarly we note that the $\nitrgn$ abundance of the $15\,\msun$ model is approximately 1 dex lower at the end of He-burning than its final value in \Cref{fig:enrichment}, again due to an increase in $\nitrgn$ through a H-He interaction in the final evolutionary stages. 

  \begin{figure}
  \centering
  \includegraphics[width=\linewidth]{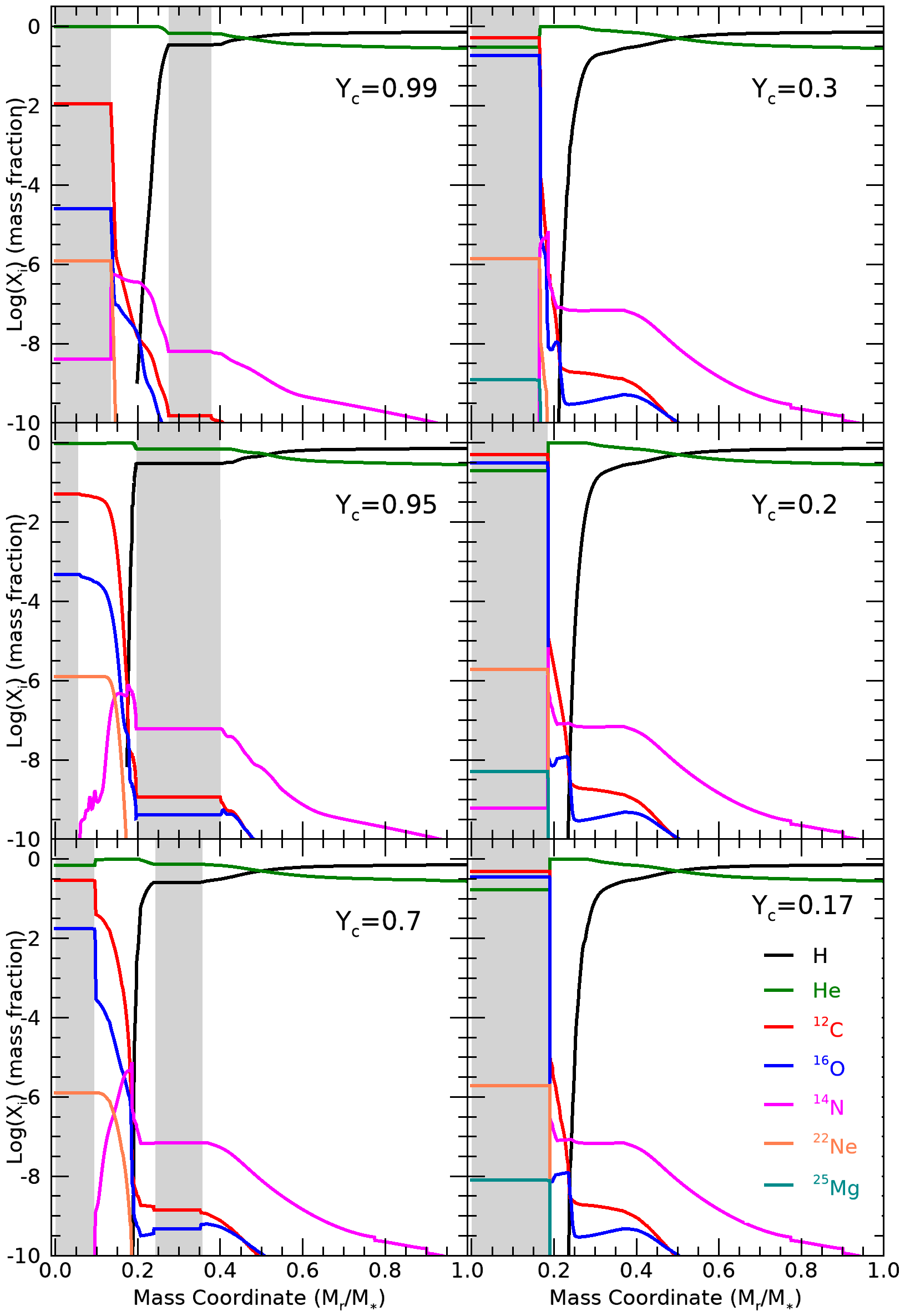}
      \caption{Evolution of the abundance profile for the $20\,\msun$ model rotating with initial velocity $\vini$=0.4$\vcrit$. Abundance profiles are shown for six points during He-burning where the central He mass fraction is 99\%, 95\%, 70\%, 30\%, 20\% and 17\%. Species are indicated by the legend.
              }
         \label{fig:20rot_abundances}
  \end{figure}
%

We find that the 12$\,\msun$ and $15\,\msun$ models experience significantly higher $\nitrgn$ production during He-burning than other models (\Cref{fig:Nevol_rot}). To understand this we can refer back to the abundance profiles in \Cref{fig:Energy_12v15} where we examined the interior structure of these models at various points of He-burning. There we see that the conditions of these models are ideal for maximising $\nitrgn$ production. Rotational mixing gradually brings CO outwards from the core during He-burning (\Cref{subfig:Abund_12v15a,subfig:Abund_12v15b,subfig:Abund_12v15c}), delivering it to the H-burning shell which is the dominant source of luminosity for the star (\Cref{subfig:Energy_12v15a,subfig:Energy_12v15b,subfig:Energy_12v15c}). Coming back to \Cref{fig:Nevol_rot}, we see that the production of $\nitrgn$ is more gradual for the $12\,\msun$ model. This is because the H shell receives less CO and the CNO cycle is not strong enough to trigger convection. The radiative nature of the shell does not affect the growing He core, as is the case for the $15\,\msun$ model (see \Cref{subsec:internalstruc}), thus allowing for continuous $\nitrgn$ enrichment. 

We observe a sudden drop in $\nitrgn$ yield for the $20\,\msun$ model, followed by a steady increase for higher masses (\Cref{fig:enrichment}). Our question then becomes, what changes between the $15\,\msun$ and $20\,\msun$ models to hinder enrichment? Through examining the interior structure of the $20\,\msun$ rotator we observe that it experiences its CNO boost at the beginning of He-burning. This causes the core to recede, which hinders the C and O that can be produced and delivered to this region where $\nitrgn$ enrichment occurs.

   \begin{figure}
   \centering
   \includegraphics[width=\linewidth]{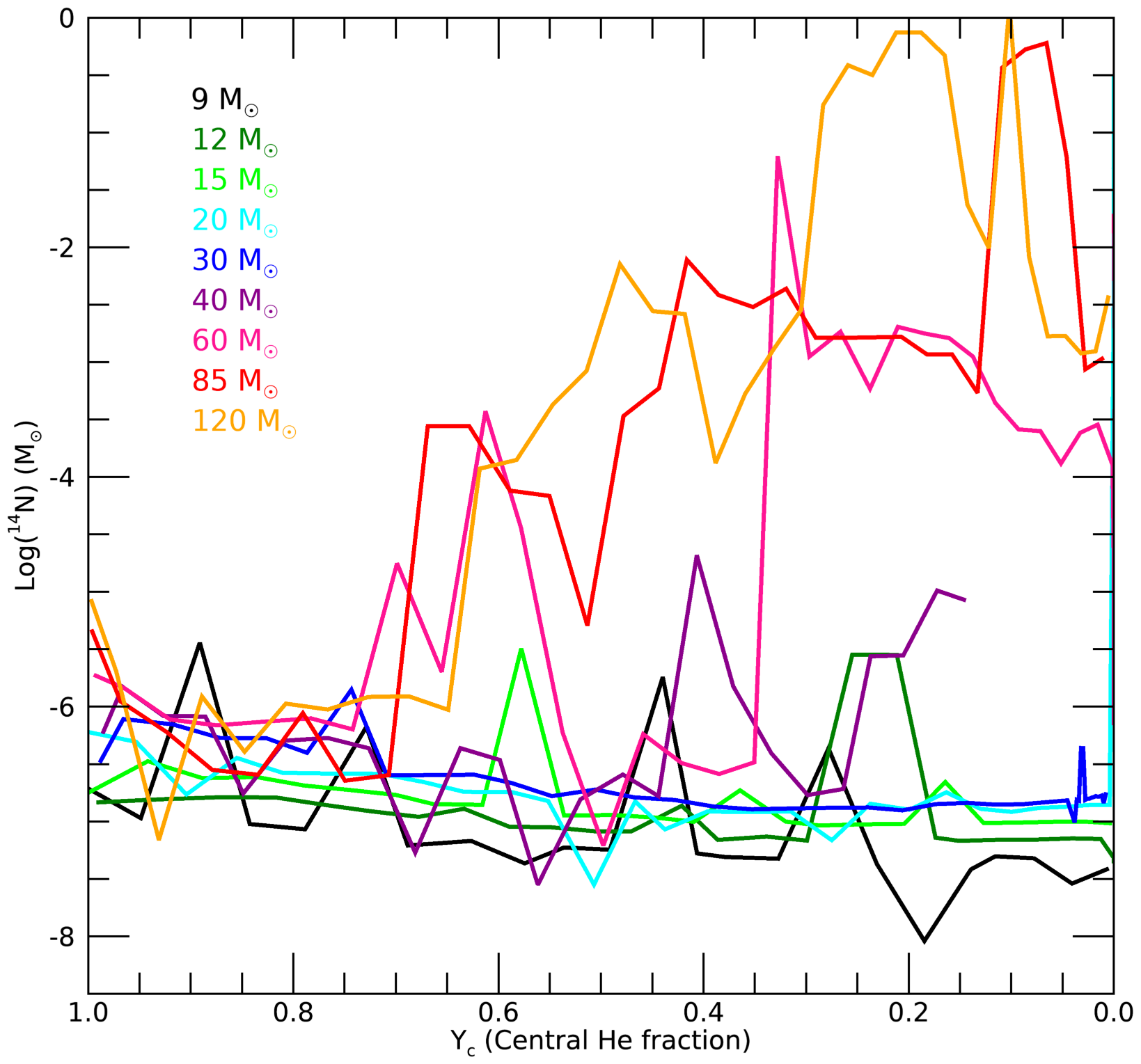}
      \caption{Same as \Cref{fig:Nevol_rot} but for non-rotating models.
              }
         \label{fig:Nevol_norot}
   \end{figure}
%

The effect of this early CNO boost on the stellar structure of the $20\,\msun$ model is shown in the three left panels of \Cref{fig:20rot_abundances}. Here we have plotted the abundance profile of the $20\,\msun$ rotator at three different stages of He-burning, before the CNO boost ($Y_{\rm c}=0.99$), shortly after the boost ($Y_{\rm c}=0.95$), and one to show the structure later on in He-burning ($Y_{\rm c}=0.7$). Between $Y_{\rm c}=0.99$ and $Y_{\rm c}=0.95$ we note the large increase in size of the convective H shell and the resulting retraction of the core. Following the CNO boost, as the star regains equilibrium, the core begins to grow again aided by rotational mixing. As the core grows the H shell moves outwards, see \Cref{sec:roteffects}, and so the CO produced in the He core never reaches the H shell. In other words, rotational mixing of material from the core is slower than the moving H shell. This is particularly evident from the bottom left panel ($Y_{\rm c}=0.7$) in \Cref{fig:20rot_abundances}, where the base of the H shell is now at a mass coordinate of $\sim\! 0.25\,\mstar$ while rotationally mixed CO extends only to $\sim\! 0.2\,\mstar$. Essentially the early triggering of the CNO boost limits the amount of CO that reaches a H-burning region, and in turn limits $\nitrgn$ enrichment.

The growing He core can have another effect however, if the He-burning core expands into a region where $\nitrgn$ has formed it converts this $\nitrgn$ into $^{22}\mathrm{Ne}$. This is what happens towards the end of He-burning for the $20\,\msun$ rotator and explains the dip in $\nitrgn$ for $Y_{\rm c}\leq0.3$ in \Cref{fig:Nevol_rot}. This effect is shown in the right panels of \Cref{fig:20rot_abundances}. At $Y_{\rm c}=0.3$ (top right panel \Cref{fig:20rot_abundances}) we note the peak in $\nitrgn$ just outside the He core. By $Y_{\rm c}=0.2$ (middle right panel \Cref{fig:20rot_abundances}) the growing core has engulfed this $\nitrgn$ rich region leading to an increase in $^{22}\mathrm{Ne}$ in the core. Finally, by $Y_{\rm c}=0.17$ (bottom right panel \Cref{fig:20rot_abundances}) the $\nitrgn$ transported to the core has been converted into $^{22}\mathrm{Ne}$ and subsequently into $^{25}\mathrm{Mg}$ through the s-process.

For models with 30-60$\,\msun$ in \Cref{fig:Nevol_rot} we see a largely constant $\nitrgn$ abundance through He-burning, owing to the early CNO boost as is the case with the $20\,\msun$ model, but without $\nitrgn$ being swallowed up by the core. In fact, for models $\geq 30\,\msun$ the CNO boost actually occurs before He-ignition. The most massive rotating models in our grid, 85$\,\msun$ and $120\,\msun$, see more significant $\nitrgn$ production during He-burning, mainly because they have larger He cores and produce more CO for enrichment. 

  \begin{figure}
  \centering
  \sbox0{\includegraphics{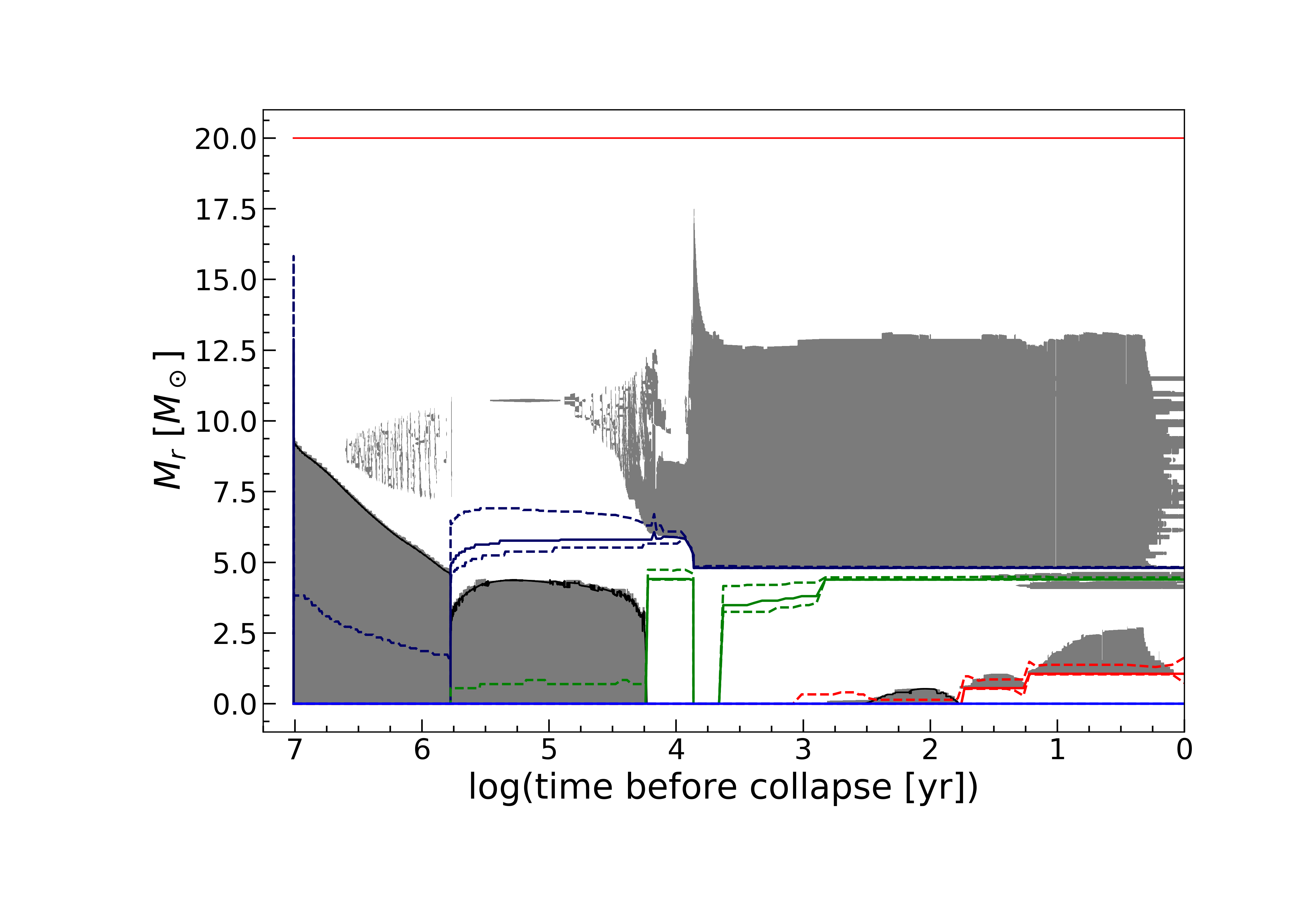}}%
  \includegraphics[clip,trim={0.1\wd0} {0.08\wd0} {0.05\wd0} {0.08\wd0}, width=\linewidth]{Kipp_P020z00S0.png}
      \caption{Kippenhahn diagram of the non-rotating $\mini=20\,\msun$ model, showing the evolution of the stellar structure in terms of the mass coordinate as a function of time to core collapse. The red line at the top indicates the total mass of the star. The grey-shaded areas correspond to convective regions. The solid (dashed) lines correspond to the peak (10\%) of the energy generation rate for H-burning (blue), He-burning (green), and C-burning (red). 
              }
         \label{fig:20norot_kipp}
  \end{figure}
%

\subsection{Non-rotating models}

Our non-rotating models do not show a clear trend of increasing $\nitrgn$ enrichment as a function of \mini\ (\Cref{fig:enrichment}), and instead display non-monotonic behavior. The key effects that influence this variety in $\nitrgn$ enrichment are: H-He shell interactions at late phases, and interaction of the He core with $\nitrgn$ which converts it to $^{22}\mathrm{Ne}$. To investigate this, we look at the time evolution of $\nitrgn$ abundance through He-burning for our non-rotating models in \Cref{fig:Nevol_norot}. We see that all non-rotating models $< \!60\,\msun$ have little $\nitrgn$ enrichment during He-burning. Without the aid of rotational mixing there is less CO available to the H shell to produce $\nitrgn$. However, as initial mass increases, the relative core size also increases, making it easier for CO to reach the H shell, and we see greater enrichment for the 60, 85 and $120\,\msun$ models in \Cref{fig:Nevol_norot}.

The largest $\nitrgn$ production seen for non-rotating models in \Cref{fig:enrichment} is that of the $20\,\msun$ non-rotator, which experiences a large CNO boost when the H shell moves inwards following He-burning. This model is therefore an example of how H-He shell interactions complicate any trends we may predict for $\nitrgn$ enrichment. This large CNO boost can be seen in \Cref{fig:20norot_kipp} where the convective region in the envelope suddenly increases reaching a mass coordinate of $\sim\!\! 17.5\,\msun$. The contraction of the star in this late phase brings the H shell inwards where He-burning had previously taken place, and the CO rich region allows for sudden and strong $\nitrgn$ production.

  \begin{figure*}
 \includegraphics[width=0.95\textwidth]{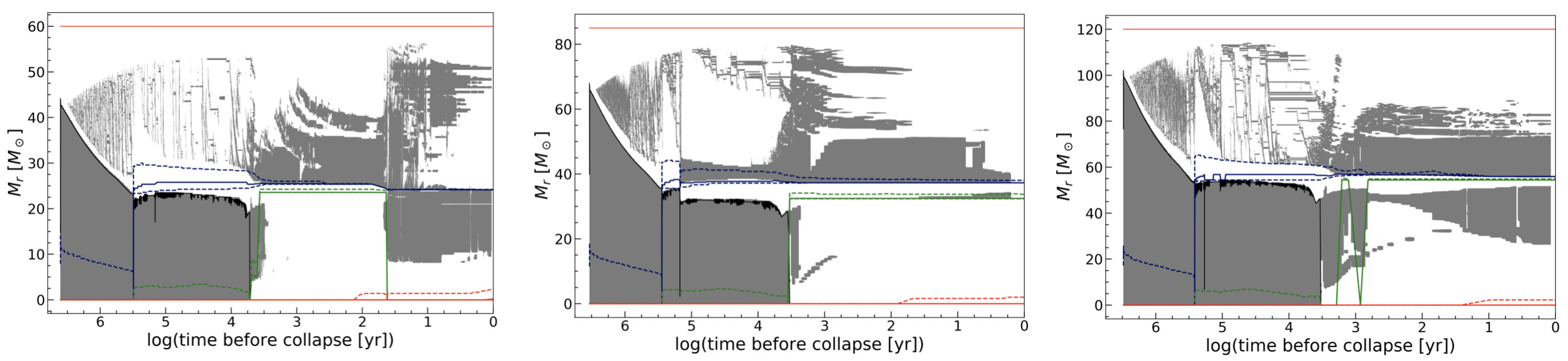}%
      \caption{Similar to Fig.~\ref{fig:20norot_kipp}, but for non-rotating models with $\mini=60\,\msun$ (left), $\mini=85\,\msun$ (middle), and $\mini=120\,\msun$ (right). 
              }
         \label{fig:60_85_120_kipp}
  \end{figure*}
%

What is also important to note about \Cref{fig:enrichment}, which can be seen in \Cref{fig:Nevol_norot}, is that for non-rotating models where $\nitrgn$ production occurs close to the He core boundary the $\nitrgn$ abundance is highly variable. We see this behaviour for the 60, 85, and 120$\,\msun$ models. The proximity of the He core to $\nitrgn$ rich regions means it can easily mix $\nitrgn$ inwards to form $^{22}\mathrm{Ne}$, which in \Cref{fig:Nevol_norot} appears as a rather jagged evolution of $\nitrgn$ abundance. In some cases the growing He core actually interacts directly with the H shell as happens with the $60\,\msun$ non-rotator (see left panel of \Cref{fig:60_85_120_kipp}), giving rise to a sudden increase in $\nitrgn$ as the H shell receives a large boost in He-burning products. Interestingly, the $85\,\msun$ non-rotator does experience a CNO boost giving rise to a convective H shell and reducing the convective He core (middle panel of \Cref{fig:60_85_120_kipp}). Conversely, the H shell in the $120\,\msun$ model without rotation remains radiative throughout He-burning (right panel of \Cref{fig:60_85_120_kipp}), allowing for a progressive increase in N abundance.

\subsection{Summary}
In summary, we find that final $\nitrgn$ abundance is highly variable due to the diverse evolutionary routes to enrichment. We do find that rotational mixing helps to mix C and O out from the core to the H-burning shell, and subsequently aids $\nitrgn$ enrichment. However, rotational mixing gives rise to earlier CNO boosts which in fact hinders overall $\nitrgn$ enrichment. Early CNO boosts cause a strong retraction of the core creating significant distance between the He core and the H shell. When the core begins to grow the H shell simultaneously moves outwards, meaning that CO mixed outwards from the core struggles to reach the H shell. This effectively limits $\nitrgn$ production to whatever was achieved during this early CNO boost. It seems that the 12$\,\msun$ and $15\,\msun$ models have the ideal conditions of enough rotational mixing to deliver sufficient CO to the H shell, without triggering the CNO boost too early. \Cref{CNOboost_table} provides a summary of the nature of the CNO boosts for each of our models, the timing of the CNO boost, whether it triggers a convective H shell during the core He-burning phase, in addition to the $\nitrgn$ yields from \Cref{gridtable2}. 

Additionally, there are multiple channels for CO elements interacting with H-burning regions that are not limited to this outward mixing during He-burning. Many of these channels are discussed in \citet{Clarkson2020} as H-He interactions, and consist of events such as the inward moving H shell following He-burning seen for the $15\,\msun$ rotator and the $20\,\msun$ non-rotator. It is clear from the discrepancy between the final abundance in \Cref{fig:Nevol_rot,fig:Nevol_norot} and the values in \Cref{fig:enrichment}, that these interactions in late burning stages can give rise to significant enrichment. Given that not all of our models evolve past the end of He-burning, further work is needed to assess the frequency of these H-He shell interactions among our models. The nature of these H-He interactions affects the production of i-process elements \citep{Clarkson2018,Clarkson2020}, and has important implications for the most metal-poor stars observed and their constraints on the first stars, so these interactions in late phases of the evolution are certainly worth further attention. However, a detailed investigation of these nuclear processes and comparison with higher metallicity models is beyond the scope of this paper and will be discussed in future work. 

   \begin{figure}
   \centering
   \includegraphics[width=0.9\linewidth]{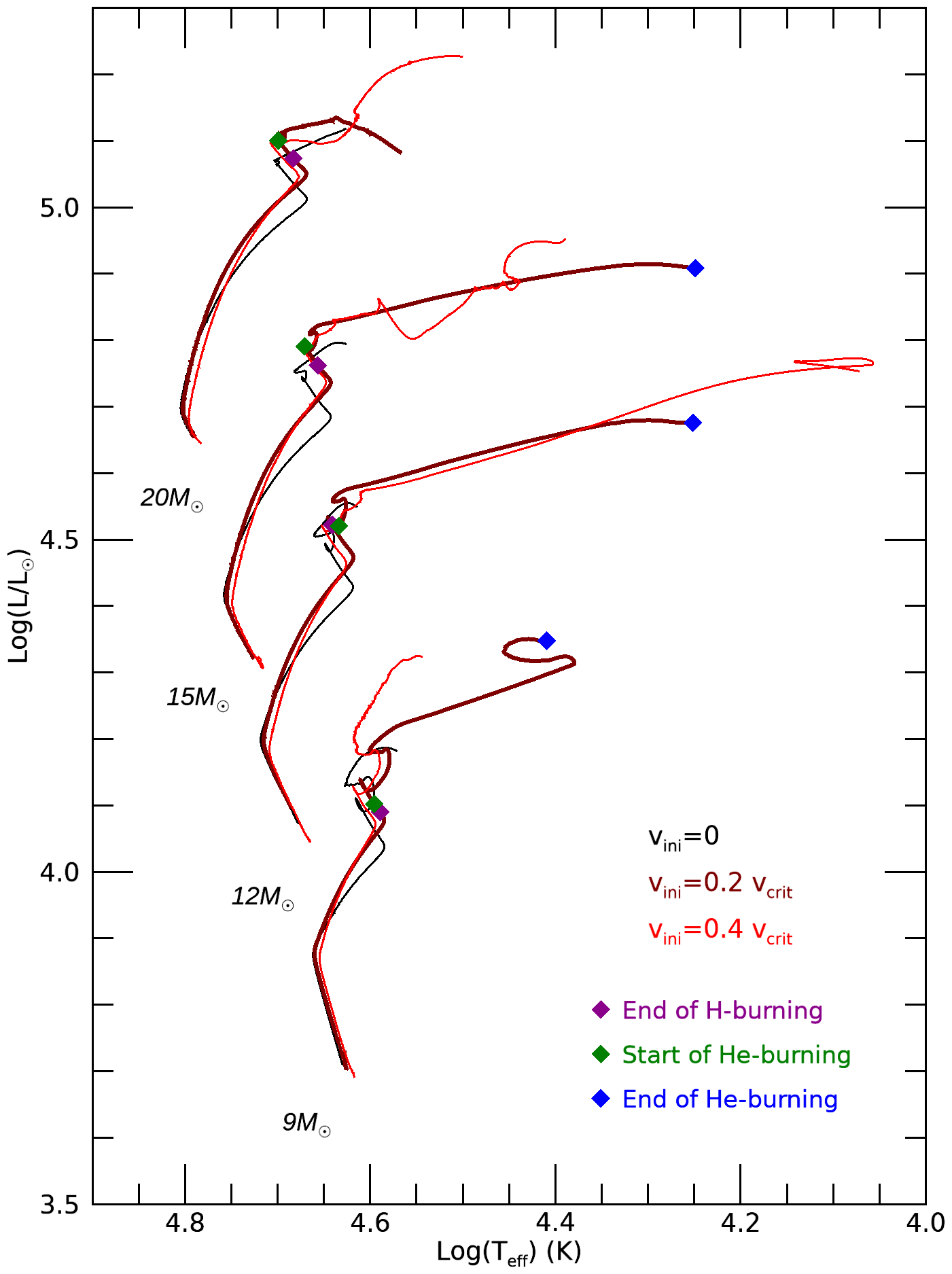}
      \caption[Evolution along the HR diagram for the three rotational velocities]{Evolution along the HR diagram for models 9-20$\,\msun$ with three rotational velocities, with our non-rotating model in black, rotating at $\vini=0.2 \, \vcrit$ in maroon, and $\vini=0.4 \, \vcrit$ in red. Models are evolved to the end of He-burning, with the exception of the $20\,\msun$ model with $\vini=0.2\,\vcrit$ which was evolved until $Y_{\rm c}=0.31$. Evolutionary phases for the $\vini=0.2 \, \vcrit$ are indicated by the legend.
              }
         \label{fig:HRD_v2v4}
   \end{figure}

\section{Varying initial rotational velocity}\label{change_vel}
In this section we will compare our models rotating at an initial velocity $\vini=0.4 \, \vcrit$ with the slower rotators in our grid of initial rotation $\vini=0.2 \, \vcrit$. As discussed in \Cref{sec:ingredients}, this will allow us to infer trends in evolutionary behaviour with rotation.

\begin{figure*}
    \centering
    \begin{subfigure}[t]{\textwidth}
        \centering
        \includegraphics[width=0.7\linewidth]{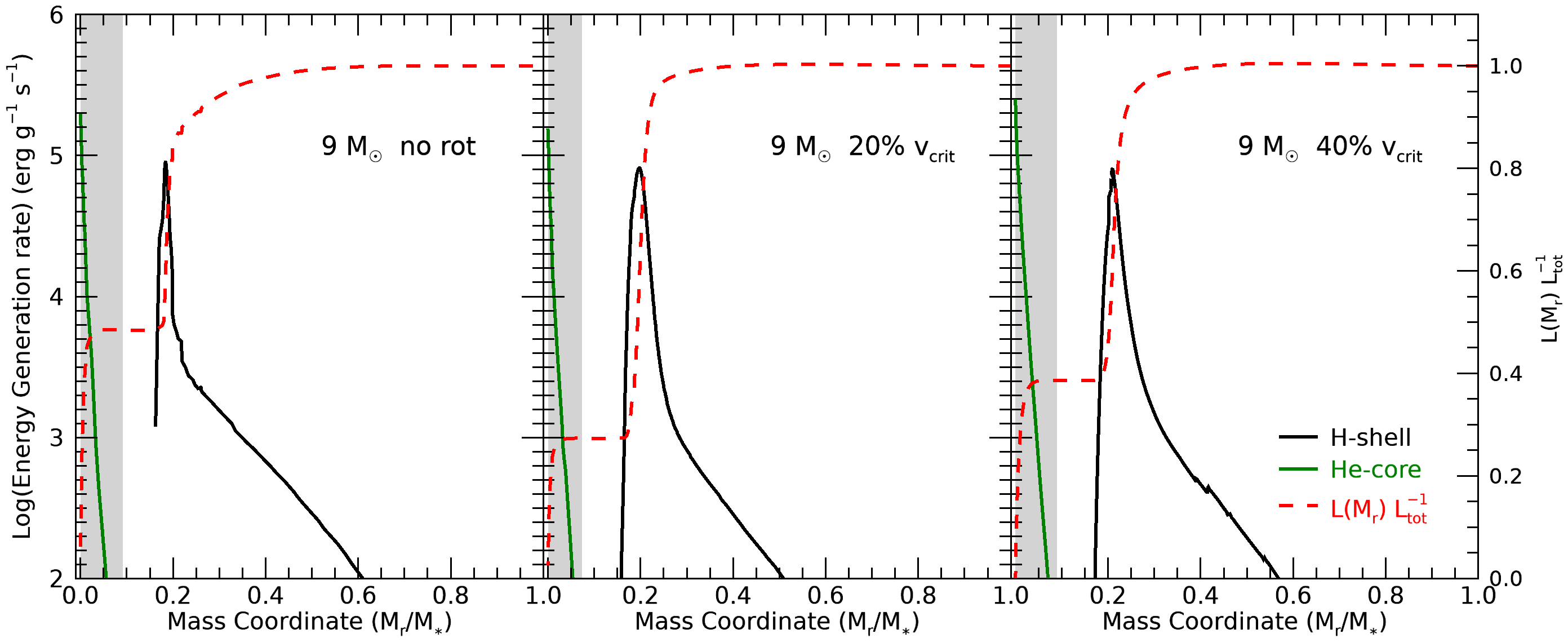}
        \caption{Energy Generation at $Y_{\rm c}$=0.5}
            \label{subfig:Energy_9a}
    \end{subfigure}\vspace{0.5cm}
    \begin{subfigure}[t]{\textwidth}
        \centering
        \includegraphics[width=0.7\linewidth]{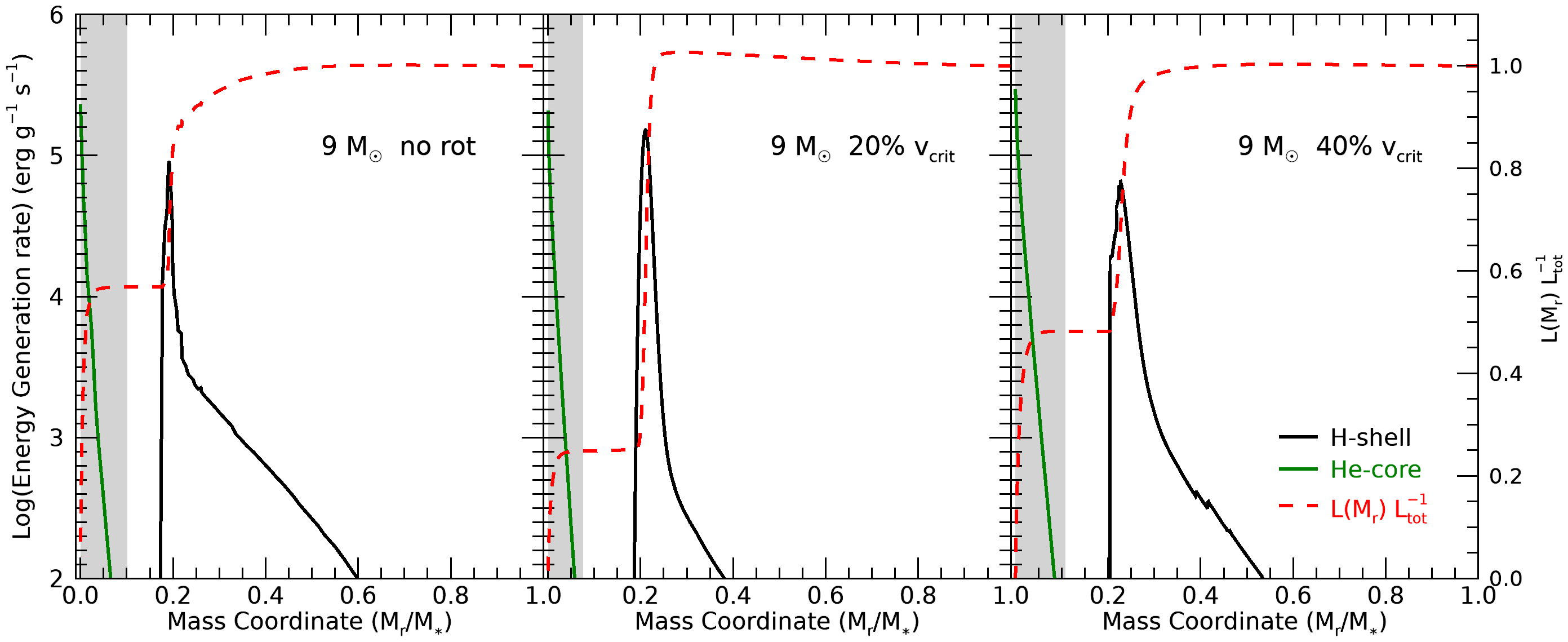}
        \caption{Energy Generation at $Y_{\rm c}$=0.25}
        \label{subfig:Energy_9b}
    \end{subfigure}
    \caption{Similar to \Cref{subfig:Energy_12v15a,subfig:Energy_12v15b,subfig:Energy_12v15c}, but for $9\,\msun$ models of rotational velocities $\vini=0, 0.2, 0.4 \, \vcrit$ at two stages during He-burning indicated by their values of $Y_{\rm c}$.}
    \label{fig:Energy_9}
\end{figure*}

From \Cref{fig:HRD_v2v4} we can see that varying the rotational velocity does lead to changes in the HR evolution of these models. There appears to be a general trend where slower rotators evolve similarly to models of lower initial mass with higher rotation. For example, the $20\,\msun$ model with $\vini\!=\!0.2 \, \vcrit$ evolves similarly to the $15\,\msun$ $\vini\!=\!0.4 \, \vcrit$ model during He-burning. That is to say that it experiences a similar CNO boost where the strong triggering of convection in the H shell leads to a retraction of the He core and subsequent decrease in luminosity, see \Cref{subsec:internalstruc}. This is evident in \Cref{fig:HRD_v2v4} from the sharp decrease in luminosity of the $20\,\msun$ $\vini\!=\!0.2\,\vcrit$ model, which resembles the beginning of the luminosity dip experienced by the $15\,\msun$ fast rotating model. Essentially, with less rotational mixing at this lower rotational velocity, less CO is delivered to the H shell so convection is triggered in the H shell for higher initial masses than would be the case at higher rotation. The 12$\,\msun$ and $15\,\msun$ models with $\vini\!=\!0.2\,\vcrit$ both maintain radiative H shells throughout He-burning and therefore see significant expansion and decrease to lower effective temperatures. The $12\,\msun$ fast rotator ($\vini\!=\!0.4 \, \vcrit$) sees greater expansion than the slower rotator because the H shell is stronger while still remaining radiative and migrating outwards (see \Cref{subsec:internalstruc}). 
The $9\,\msun$ model shows particularly interesting behaviour as rotational velocity increases. The differences in energy generation for the three $9\,\msun$ models at different rotational velocities are shown in \Cref{fig:Energy_9} at different stages of He-burning. For the $\vini\!=\!0.2 \, \vcrit$ model the effect of rotation is as we see for the 12$\,\msun$ and $15\,\msun$ models where rotational mixing strengthens the H shell. However, as we move to higher rotation again, $\vini=0.4 \, \vcrit$, the H shell is in fact weaker. We do note that the He core is larger for this higher rotational velocity. Indeed it is clear through the contribution to the total luminosity that the faster rotator is more dependent on the He core than in the slower rotator. This explains the behaviour we see on the HR diagram in \Cref{fig:HRD_v2v4}. The slow rotator evolves to low effective temperatures due to the dominant H shell, while the fast rotator remains at higher effective temperature. This emphasises the conclusion from \Cref{sec:roteffects}, that the relative core and shell strength dominate the evolution of the surface properties. Rotation plays a vital role in this balance of core and shell strength, it provides additional fuel through rotational mixing into the core, but it also boosts the H shell through delivery of heavy elements from the core which can result in core retraction. It seems then that there are two competing effects, rotational mixing either increases He core size, or it leads to a stronger dominant H shell which suppresses the core. In the $12\,\msun$ case the H shell wins out, while rotational mixing favours growth of the core for the $9\,\msun$ model. 

\section{Concluding remarks}\label{conclusions}
Pop~III models are unique in their evolution in a number of ways that impact their observable features and rotational effects. Their zero-metallicity nature means that they are unable to burn Hydrogen through the CNO cycle initially and without this crucial energy supply they experience a contraction phase during the early MS. The lack of CNO elements is not only an issue for the central regions of the star but also leads to sharp $\mu$-gradients and energy increases (due to triggering of the CNO-cycle) in the stellar envelope as He-burning products are transported outwards. Therefore rotational mixing has a unique impact in these stars. We have carried out a detailed investigation of the interior structure of these models throughout their evolution and how this has driven the evolution of the surface properties of these stars. This has given us a new understanding and perspective on the role of rotation for Pop~III stars. The following are our main conclusions from this work.
\begin{itemize}
    \item Rotation has a significant impact on the observable signatures of Pop~III stars through two main effects. Firstly, rotational mixing brings additional fuel into the nuclear burning core which increases luminosity as well as stellar lifetimes. Secondly, rotational mixing brings He-burning products from the core to the H-burning shell during later evolutionary phases, which changes the temperature profile, and can lead to significant expansion in some models depending on the relative core size. The relative core size is crucial here, because the contribution of the shell and the core to the total energy produced tells us about the structure of the star and what dominates with regard to the evolution of the surface properties.
    \item Despite weaker meridional currents for Pop~III stars angular momentum can build up at the surface for fast rotating massive models because of their negligible mass loss through radiative winds. This spin up brings models $\mini \geq 60\,\msun$ with $\vini=0.4 \, \vcrit$ to critical rotation on the MS which leads to increased mass loss with as much as 3.5$\,\msun$ of material lost for our most massive model of $\mini\!=\!120\,\msun$. Further work is needed to determine the nature of this mass loss.
    \item Rotational mixing strongly affects metal enrichment, but does not always increase metal production as we see at higher metallicities. Rotation leads to an earlier CNO boost to the H shell during He-burning, which may hinder metal enrichment. This is true for precise mass and initial velocity domains, and only for the core He-burning phase. In these cases the triggering of convection by the CNO boost in the H shell causes a retraction of the He-burning core. As the core grows the H shell moves outwards and does so more quickly than He-burning products can be rotationally mixed out from the core, therefore hindering the interaction of these products with the H-burning shell, which is required for metal enrichment. The H-He shell interactions after core He-burning play a crucial role in metal production, and there rotation may boost enrichment. This highlights the complexity in the metal enrichment processes of these models. A detailed understanding of the interior structure is therefore required to accurately predict metal yields.  
    \item Through comparing our models with slower rotators at $\vini=0.2 \, \vcrit$, we have shown that a general trend exists where higher rotation in a model of a certain initial mass leads to similar evolutionary behaviour of a more massive model with lower initial rotation. There is a trade off between increasing initial mass and rotational velocity in order to see the same evolution of model structure. For example, the $\mini\!=\!20\,\msun$ model with $\vini\!=\!0.2 \, \vcrit$ sees a strong CNO boost with significant change to total luminosity, similarly to the $\mini\!=\!15\,\msun$ model with $\vini\!=\!0.4 \, \vcrit$ model which also sees this behaviour, indicated by the luminosity dip on the HR diagram.
\end{itemize}


\section*{Acknowledgements} We wish to acknowledge the Irish Research Council for funding this research, as well as "ChETEC" COST Action (CA16117), supported by COST (European Cooperation in Science and Technology) which aided our collaboration with co-authors.

\section*{Data availability} The derived data generated in this research will be shared on reasonable request to the corresponding author.

\bibliography{refpapers}
\bibliographystyle{mnras}


\begin{landscape}
\begin{table}
 \centering   
 \caption{Summary of our GENEC model grid for Pop~III stars. We show the initial mass (column 1), initial ratio between surface rotational velocity and critical velocity (column 2), surface equatorial velocity (column 3), and the  properties of our models at end H-burning (columns 4--8) and end He-burning (columns 9--13). MS and He-burning lifetimes are given by $\tau_{\rm H}$ and $\tau_{\rm He}$ respectively. For each phase we quote $\upsilon_{\rm eq}$ (the velocity at the equator) and $Y_{\rm surf}$ (He mass fraction at the surface). }
\begin{tabular}{|c|c|c|c|c|c|c|c|c|c|c|c|c}
   \hline
  \multicolumn{3}{c|}{} & 
  \multicolumn{5}{c|}{End MS} & 
  \multicolumn{5}{c}{End He-burning} \\
    \hline
    \noalign{\smallskip}
    Initial Mass (M$_\odot$)  &  $\vini/\vcrit$ & $\upsilon_{\rm eq}$ (km/s) & $\tau_{\rm H}$ (yrs) & M (M$_\odot$) & $\upsilon_{\rm eq}$ (km/s) & $\upsilon_{\rm eq}/\vcrit$ & $Y_{\rm surf}$ & $\tau_{\rm He}$ (yrs) & M (M$_\odot$) & $\upsilon_{\rm eq}$ (km/s) & $\upsilon_{\rm eq}/\vcrit$ & $Y_{\rm surf}$\\
    \noalign{\smallskip}

    \hline
    \noalign{\smallskip} 
9  &       0  &      0  &  1.77097e+07  &      9  &      0  &         0  &     0.248359  &  1.90800e+06  &      9  &      0  &         0  &     0.248359\\
12  &       0  &      0  &  1.78575e+07  &      12  &      0  &      0  &     0.248360  &  1.07950e+06  &      12  &      0  &      0  &     0.248360\\
15  &       0  &      0  &  1.29865e+07  &      15  &      0  &      0  &     0.248364  &  793225  &      15  &      0  &      0  &     0.248364\\
20  &       0  &      0  &  9.50960e+06  &      20  &      0  &         0  &     0.248372  &  573432  &      20  &      0  &         0  &     0.248372\\
30  &       0  &      0  &  6.16050e+06  &      30  &      0  &         0  &     0.248378  &  419249  &      30  &      0  &         0  &     0.248378\\
40  &       0  &      0  &  4.79495e+06  &      40  &      0  &         0  &     0.248380  &  356070  &      40  &      0  &         0  &     0.248380\\
60  &       0  &      0  &  3.66825e+06  &      60  &      0  &         0  &     0.248382  &  302693  &      60  &      0  &         0  &     0.248382\\
85  &       0  &      0  &  3.07592e+06  &      85  &      0  &      0  &     0.248383  &  271411  &      85  &      0  &      0  &     0.248388\\
120  &       0  &      0  &  2.73616e+06  &      120  &      0  &      0  &     0.248383  &  253424  &      120  &      0  &      0  &     0.248383\\
    \noalign{\smallskip}
    \hline
    \noalign{\smallskip}
9  &     0.4  &      372  &  2.09097e+07  &      9  &      274  &     0.379501  &     0.285829  &  2.45972e+06  &
      9  &      60.3  &     0.111874  &     0.286148\\
12  &     0.4  &      371  &  1.97218e+07  &      12  &      292  &     0.407252  &     0.286611  &  2.41656e+06  &
      12  &      31.6  &     0.189222  &     0.312177\\
15  &     0.4  &      427  &  1.49696e+07  &      15  &      289  &     0.400832  &     0.281065  &  1.23126e+06  &
      15  &      67.5  &     0.191761  &     0.287834\\
20  &     0.4  &      526  &  1.07123e+07  &      20  &      309  &     0.413102  &     0.276504  &      971262  &
      20  &      192  &     0.435374  &     0.276718\\
30  &     0.4  &      527  &  6.99825e+06  &      30  &      359  &     0.472368  &     0.276930  &      498099  &
      30  &      297  &     0.622642  &     0.278309\\
40  &     0.4  &      562  &  5.33301e+06  &      40  &      445  &     0.573454  &     0.267699  &      429791  &
      40  &      340  &     0.514372  &     0.268766\\
60  &     0.4  &      613  &  4.06187e+06  &      59.8  &      657  &     0.872510  &     0.264842  &  360821  &      59.7  &      302  &     0.749380  &
     0.268879\\
85  &     0.4  &      659  &  3.29659e+06  &      84.1  &      551  &     0.703704  &     0.265406  &  309186  &      84.0  &      269  &     0.734973  &
     0.268855\\
120  &     0.4  &      708  &  2.88846e+06  &      116.5  &      473  &     0.639189  &     0.302554  &  254127  &      116.5  &      274  &     0.591793  &
     0.311393\\
    \noalign{\smallskip}
    \hline    
    \end{tabular}
    \label{gridtable1}

 \end{table}
 \end{landscape}


\begin{table*}
 \centering   
 \caption{Summary of the final properties of our model grid. We show the initial mass, rotational velocity as a fraction of critical velocity, evolutionary stage reached by end of model run, as well as He core and CO core masses, and the total mass of \nitrgn, $^{12}\mathrm{C}$, and $^{16}\mathrm{O}$ above the gravitational remnant mass (see \Cref{sec:enrichment}) at the evolutionary stage given in the third column.}
\begin{tabular}{|c|c|c|c|c|c|c|c|}
   
    \hline
    \noalign{\smallskip}
    Initial Mass (M$_\odot$)  &  $\vini/\vcrit$ & evolutionary stage & M$_{\mathrm{He}}$ (M$_\odot$) & M$_{\mathrm{CO}}$ (M$_\odot$) & N$_{\mathrm{prod}}$ (M$_\odot$) & C$_{\mathrm{prod}}$ (M$_\odot$) & O$_{\mathrm{prod}}$ (M$_\odot$)\\
    \noalign{\smallskip}
    \hline
    \noalign{\smallskip} 
    9   &  0  &  degenerate before C-ignition & 1.857 & 1.039 & 2.25$\times 10^{-8}$ & 9.51$\times 10^{-2}$ & 2.07$\times 10^{-2}$   \\
    12  &  0  &  End He-burning & 2.700 & 1.819 & 1.21$\times 10^{-7}$ & 0.6096 & 0.6547\\
    15  &  0  &  End He-burning & 4.194 & 2.551 & 1.75$\times 10^{-7}$ & 0.8802 & 0.1.3822\\
    20  &  0  &  Ne-burning & 4.801 & 4.386 & 1.39$\times 10^{-3}$ & 0.6441 & 1.6452\\
    30  &  0  &  He-burning, $Y_\mathrm{c}$=0.007 & 11.813 & 8.586 & 3.54$\times 10^{-7}$ & 3.2524 & 8.8982 \\
    40  &  0  &  Ne-burning & 14.996 & 13.202 & 1.91$\times 10^{-7}$ & 2.1055 & 7.8767\\ 
    60  &  0  &  C-burning  & 24.135 & 24.046 & 1.16$\times 10^{-4}$ & 3.043 & 15.697 \\
    85  &  0  &  C-burning  & 52.225 & 32.389 & 4.63$\times 10^{-5}$ & 9.8349 & 47.1081 \\
    120 &  0  &  End He-burning  & 73.277 & 54.406 & 1.38$\times 10^{-5}$ & 4.8774 & 40.9261 \\ 
    \noalign{\smallskip}
    \hline
    \noalign{\smallskip}
    9   &  0.4  &  degenerate before C-ignition & 2.328 & 1.336 & 2.13$\times 10^{-3}$ & 0.2326 & 0.2793\\
    12  &  0.4  &  Ne-burning & 3.952 & 2.355 & 8$\times 10^{-3}$ & 0.3127 & 0.9704\\
    15  &  0.4  &  C-burning & 2.852 & 2.266 & 0.0123 & 0.4366 & 0.6262\\
    20  &  0.4  &  C-burning & 7.198 & 4.297 & 4.94$\times 10^{-7}$ & 1.2338 & 4.1089\\
    30  &  0.4  &  End He-burning & 9.82 & 6.703 & 1.4$\times 10^{-6}$ & 0.8876 & 3.6803\\
    40  &  0.4  &  He-burning, $Y_\mathrm{c}$=0.04 & 20.354 & 10.307 & 2.74$\times 10^{-6}$ & 2.3408 & 4.878\\
    60  &  0.4  &  He-burning, $Y_\mathrm{c}$=0.002 & 35.122 & 20.936 & 1.08$\times 10^{-5}$ & 5.6166 & 27.2175\\
    85  &  0.4  &  He-burning, $Y_\mathrm{c}$=0.027 & 52.364 & 31.286 & 1.73$\times 10^{-5}$ & 10.9537 & 41.6622 \\
    120 &  0.4  &  He-burning, $Y_\mathrm{c}$=0.092 & 73.938 & 56.399 & 1.42$\times 10^{-3}$ & 28.3493 & 65.812\\
    \noalign{\smallskip}
    \hline    
    \end{tabular}
    \label{gridtable2}

 \end{table*}


\begin{table*}
 \centering   
 \caption{Summary of the nature of CNO boost and H shell for each model. The third column gives central He fraction at the beginning of the CNO boost when \nitrgn abundance is increased in H shell. The fourth column states whether convection is triggered in the H shell during He-burning. We note that models where H convective shells develop after He-burning are not marked here, and models which develop small temporary convective zones are not marked as having a convective shell. This typically occurs for massive non-rotators (see \citealt{Heger2000} and \citealt{Hirschi2004}). }
\begin{tabular}{c|c|c|c|c}
   
    \hline
    \noalign{\smallskip}
    Initial Mass (M$_\odot$)  &  $\vini/\vcrit$ & Stage of CNO boost ($\simeq$) & Convective H shell (He-burning) & N$_{\mathrm{prod}}$ (M$_\odot$) \\
    \noalign{\smallskip}
    \hline
    \noalign{\smallskip} 
    9   &  0  &  N/A &  & 2.25$\times 10^{-8}$ \\
    12  &  0  &  N/A &  & 1.21$\times 10^{-7}$ \\
    15  &  0  &  N/A &  & 1.75$\times 10^{-7}$ \\
    20  &  0  &  post He-burning &  & 1.39$\times 10^{-3}$ \\
    30  &  0  &  N/A &  & 3.54$\times 10^{-7}$ \\
    40  &  0  &  N/A &  & 1.91$\times 10^{-7}$ \\ 
    60  &  0  &  $Y_\mathrm{c}$=0.75  &  & 1.16$\times 10^{-4}$ \\
    85  &  0  &  $Y_\mathrm{c}$=0.7  & \Checkmark  & 4.63$\times 10^{-5}$\\
    120 &  0  &  $Y_\mathrm{c}$=0.6  &  & 1.38$\times 10^{-5}$ \\
    \noalign{\smallskip}
    \hline
    \noalign{\smallskip}
    9   &  0.4  &  $Y_\mathrm{c}$=0.8 &  & 2.13$\times 10^{-3}$  \\
    12  &  0.4  &  $Y_\mathrm{c}$=0.7 &  & 8$\times 10^{-3}$ \\
    15  &  0.4  &  $Y_\mathrm{c}$=0.5 & \Checkmark & 0.0123 \\
    20  &  0.4  &  $Y_\mathrm{c}$=0.9 & \Checkmark  & 4.94$\times 10^{-7}$ \\
    30  &  0.4  &  $Y_\mathrm{c}$=0.95 & \Checkmark  & 1.4$\times 10^{-6}$ \\
    40  &  0.4  &  pre He-burning, $Y_\mathrm{c}$=1 & \Checkmark  & 2.74$\times 10^{-6}$ \\
    60  &  0.4  &  pre He-burning, $Y_\mathrm{c}$=1 & \Checkmark  & 1.08$\times 10^{-5}$ \\
    85  &  0.4  &  pre He-burning, $Y_\mathrm{c}$=1 & \Checkmark  & 1.73$\times 10^{-5}$ \\
    120 &  0.4  &  pre He-burning, $Y_\mathrm{c}$=1 & \Checkmark  & 1.42$\times 10^{-3}$ \\
    \noalign{\smallskip}
    \hline    
    \end{tabular}
    \label{CNOboost_table}

 \end{table*}


\appendix

\section{Onset of He-burning}
\label{sec:app1}

\begin{figure}
    \centering
    \includegraphics[width=\linewidth]{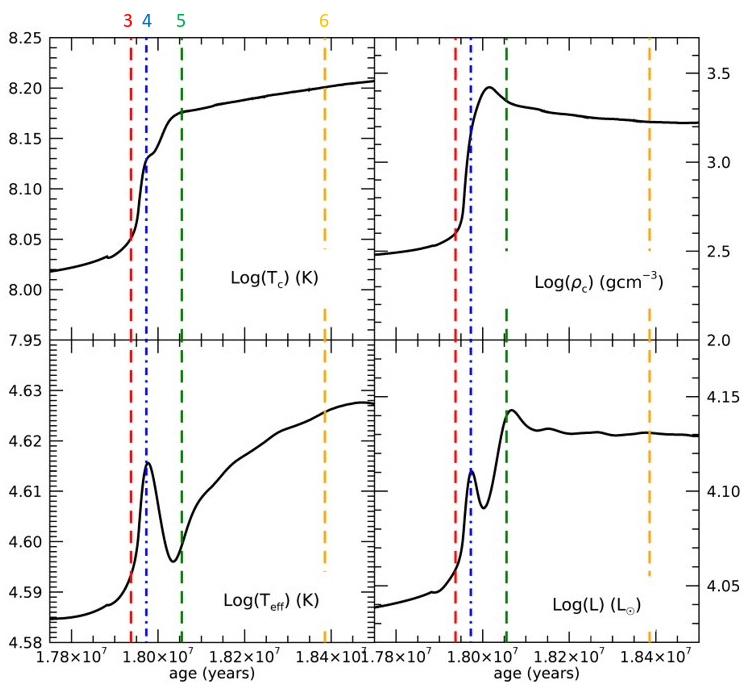}
    \caption[Central and Surface Properties for the $9\,\msun$ model]{Central and surface properties of the non-rotating $9\,\msun$ model, including central temperature ($T_c$), central density ($\rho_c$), effective temperature ($\teff$), and luminosity ($L$). Indicated by the vertical dashed lines are the ages which correspond to stages 3-6 in \Cref{fig:9HR}.}
    \label{fig:9centsurf}
\end{figure}

\begin{figure}
    \centering
    \includegraphics[width=\linewidth]{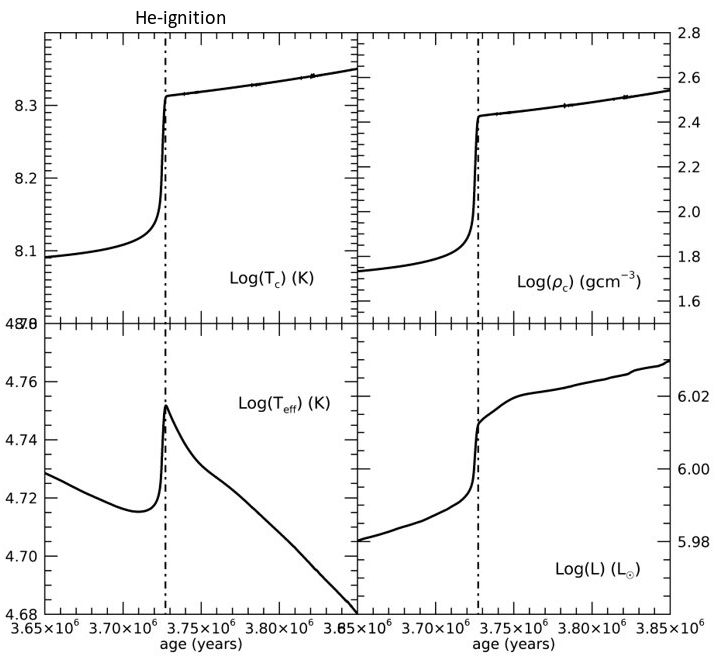}
    \caption[Central and Surface Properties for the $60\,\msun$ model]{Central and surface properties of the non-rotating $60\,\msun$ model, similarly to \Cref{fig:9centsurf}. The dashed line indicates He-ignition.}%
    \label{fig:60centsurf}
\end{figure}

\begin{figure}
    \centering
    \includegraphics[width=\linewidth]{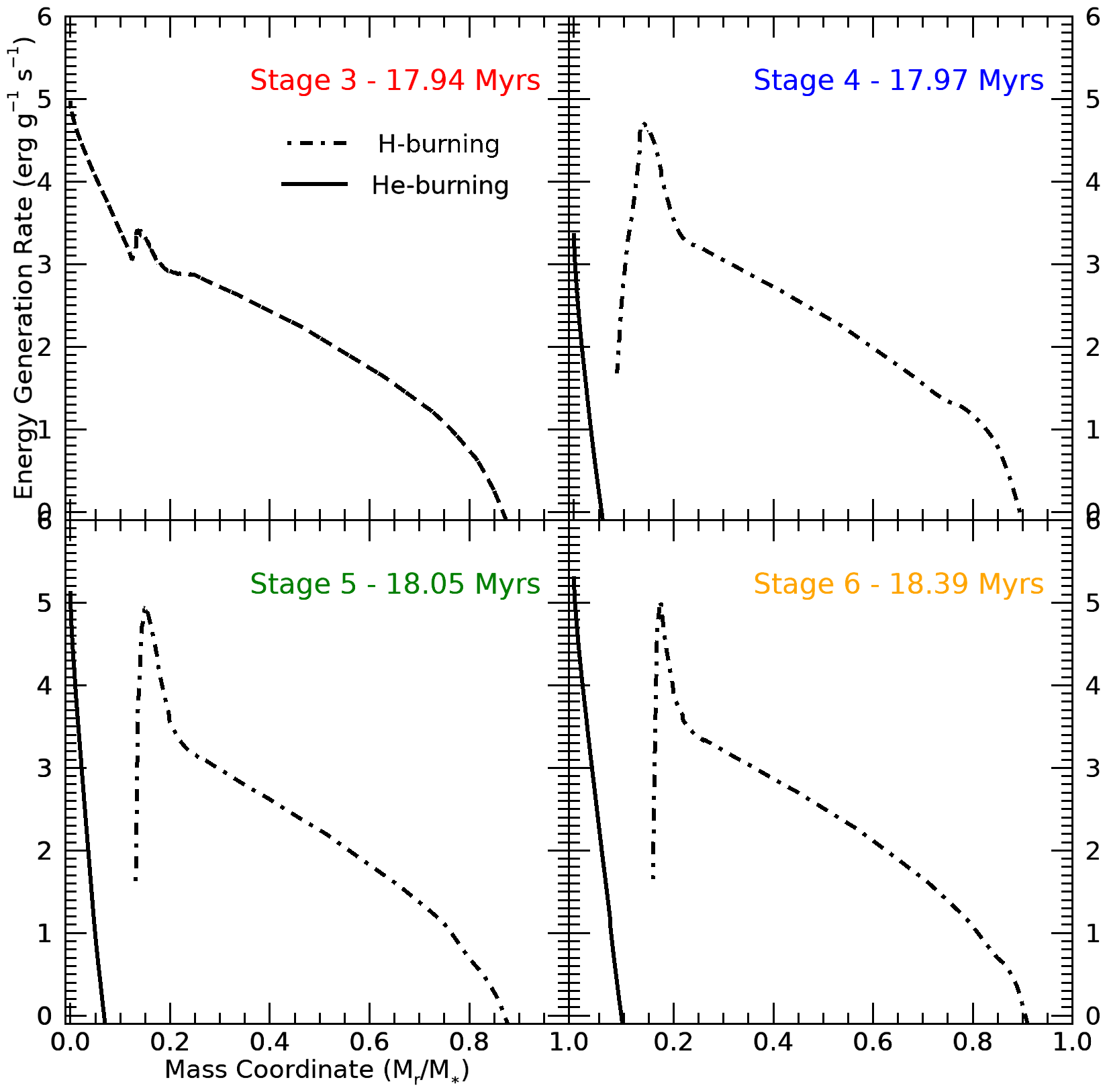}
    \caption[Energy generation rates for $9\,\msun$ model between MS and He-burning]{Energy contribution of H-burning and He-burning for the $9\,\msun$ model at stages 3-6 in \Cref{fig:9HR}. Mass coordinate illustrates how much mass lies within that region of the stellar interior with stellar centre at $M_{\rm r}$/M$_\ast=0$. }
    \label{fig:9engen}
\end{figure}

  \begin{figure}
  \centering
  \includegraphics[width=\linewidth]{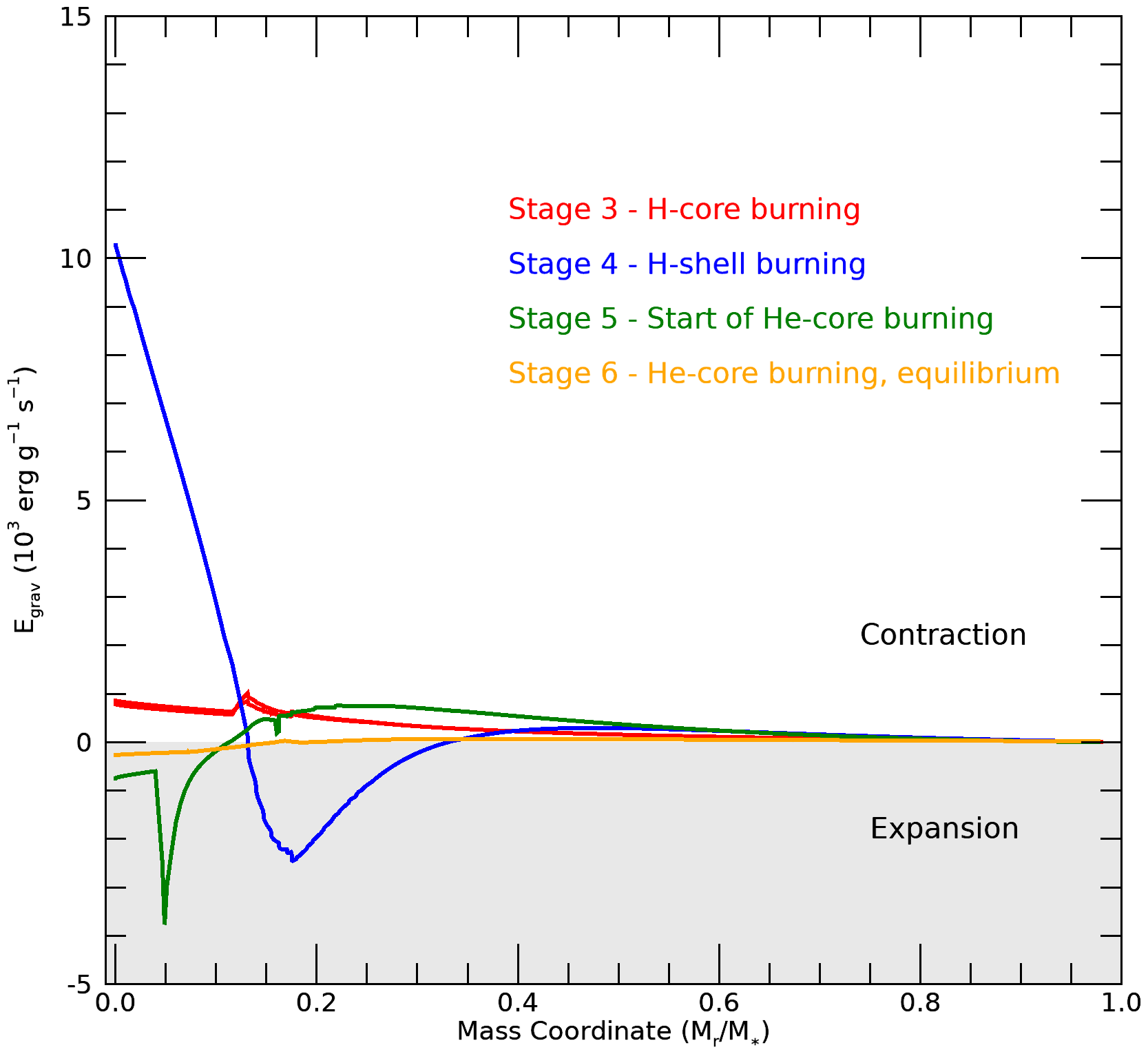}
      \caption[Gravitational energy contribution for $9\,\msun$ model]{Gravitational energy generation rates for the $9\,\msun$ non-rotating model, where positive values indicate contraction and negative values indicate expansion. Evolutionary stages 3-6 are indicated by the legend.}
         \label{fig:9egrav}
  \end{figure}


In \Cref{sec:roteffects} we discuss the distinctive feature at the onset of He-burning for lower mass models, using the non-rotating $9\,\msun$ model to illustrate the evolutionary behaviour at this stage. \Cref{fig:9HR} shows the key stages in the evolution (stages 3-6) that were used to study this prominent effect on the surface properties. To understand the behaviour of the star during this period, the surface properties were compared with the central properties, \Cref{fig:9centsurf}, for stages 3-6 in the evolution of the loop. These points correspond to significant differences in the internal structure at this stage, see \Cref{fig:9engen}. From \Cref{fig:9centsurf} we see that, while the central properties are similar to what we expect for the onset of He core burning, the surface properties reveal that more complex behaviour is going on. This is illustrated by the comparison of the $60\,\msun$ model properties (\Cref{fig:60centsurf}) where the surface temperature, bottom left panel, increases sharply as the star contracts following H core exhaustion then gradually decreases following He-ignition as the stellar envelope expands during He core burning. 

By contrast the $9\,\msun$ model shows a sharp decrease in surface temperature immediately after He-ignition followed by a gradual increase in surface temperature. Since the central conditions of the star cannot illustrate why the surface is behaving in this manner the internal structure of the star needed to be investigated. \Cref{fig:9engen} shows, for the four ages noted in \Cref{fig:9centsurf} (stages 3-6), where a H-burning shell has formed following H core exhaustion. This shell would have developed towards the end of H core burning as the star is contracting. When Hydrogen is depleted in the core the continuing contraction of the star ignites this H shell leading to a boost in luminosity at the surface, seen as the first luminosity bump in the bottom right panel of \Cref{fig:9centsurf}. He-burning then begins in the core and we get a further boost to the luminosity. However, this still does not explain the effect on surface temperature. The surprising increase in surface temperature suggests that the stellar envelope may be contracting when He core burning begins, to investigate this theory the gravitational energy contribution for the stellar interior structure was studied, plotted here as \Cref{fig:9egrav}. When positive this indicates contraction, conversely negative values indicate expansion. This therefore allows us to visualise how the stellar core and envelope react to the varying nuclear burning conditions. What can be seen from \Cref{fig:9egrav}, is that just prior to He core ignition when the H-burning shell dominates (stage 4), the now inactive core is strongly contracting, while the envelope expands due to the energy boost from this H shell. This causes the star to lose its thermal equilibrium as the connection between core and envelope becomes unstable. At 18.05 Myrs into the star's evolution (stage 5) He core burning has now begun and the star starts to regain this thermal equilibrium. As seen from \Cref{fig:9egrav} this leads to a sharp expansion of the outer core and a contraction of the envelope. The contraction of the envelope gives the surface temperature increase that we see in \Cref{fig:9centsurf}, until approximately 350'000 years later (stage 6) when the star is once again stable and can continue He-burning as seen for higher mass models such as the $60\,\msun$ example.

\label{lastpage}
\end{document}